\documentclass{aa}  

\usepackage{txfonts}
\usepackage{color}

\newcommand{\add}[1]{\textcolor{blue}{#1}}
\newcommand{\del}[1]{\textcolor{red}{{\iffalse{#1}\fi}}}
\usepackage{graphicx}

\usepackage{txfonts}

\begin{document}

   \title{Detectability of compact intermediate-mass black hole binaries as low-frequency gravitational wave sources: the influence of dynamical friction of dark matter}
     \author{Wen-Qing Jia\inst{1,2}
        \and
         Wen-Cong Chen\inst{3,4} }

   \institute{National Astronomical Observatories, Chinese Academy of Sciences, 20A Datun Road, Chaoyang District, Beijing 100101, People's Republic of China\\
         \and
         School of Astronomy and Space Sciences, University of Chinese Academy of Sciences, Beijing 100049, People's Republic of China\\
         \and
         School of Science, Qingdao University of Technology, Qingdao 266525, People's Republic of China \email{chenwc@pku.edu.cn}\\
         \and
        School of Physics and Electrical Information, Shangqiu Normal University, Shangqiu 476000, People's Republic of China }

\date{Received ; accepted }

 \abstract
{The black hole (BH) spin could significantly change the density of dark matter (DM) in its vicinity, creating a mini-spike of the density of DM. The dynamical friction (DF) between DM and the companion star of a BH can provide an efficient loss of angular momentum, driving the BH-main sequence (MS) star binary to evolve toward a compact orbit system.}
{We investigate the influence of the DF of DM on the detectability of intermediate-mass black hole (IMBH)-MS binaries as low-frequency gravitational wave (GW) sources.}
{Taking into account the DF of DM, we employ the detailed binary evolution code \texttt{MESA} to model the evolution of a large number of IMBH-MS binaries. }
{ Our simulation shows that the DF of DM can drive those IMBH-MS binaries to evolve toward low-frequency GW sources for a low donor-star mass, a high spike index, or a short initial orbital period. When the spike index $\gamma=1.60$, those IMBH-MS binaries with donor-star masses of $1.0-3.4~ M_{\odot}$ and initial orbital periods of $0.65-16.82~ \rm days$ could potentially evolve into visible LISA sources within a distance of $10~\rm kpc$.}
{ The DF of DM can enlarge the initial parameter space and prolong the bifurcation periods. In the low-frequency GW source stage, the X-ray luminosities of those IMBH X-ray binaries are $\sim 10^{35}-10^{36}~\rm erg\,s^{-1}$, hence they are ideal multimessenger objects.}
\keywords{X-ray binaries, Black holes, Dark matter, Stellar evolution, gravitational wave }

\titlerunning{Detectability of Compact IMBH Binaries as low-frequency GW Sources}
\maketitle

\section{Introduction}
Stellar-mass black holes (BHs with a mass of $\sim10-100~M_\odot$) are the evolutionary products of massive stars. Supermassive BHs (with a mass of $\sim10^{6}-10^{9}~M_\odot$) are generally located at the center of a galaxy and are thought to be the energetic central engines of an active galactic nucleus. In the mass gap ($\sim10^{2}-10^{5}~M_\odot$) between stellar-mass BHs and supermassive BHs, there may exist a transitional population called intermediate-mass black holes (IMBHs). In theory, IMBHs may form by the direct collapse from very massive Population III stars \citep{mada01},  runaway stellar collisions in dense star clusters \citep{port02,port04}, gradual growth via stellar-mass BH mergers \citep{olea06}, and an accretion onto preexisting seed BHs at the cluster center \citep{vesp10}.

IMBHs were invoked to explain ultraluminous X-ray sources (ULXs) with isotropic X-ray luminosities exceeding $10^{39}~\rm erg\,s^{-1}$ \citep{colb99,feng11}. Some works proposed that IMBHs power the ULX source in MGG-11 \citep{kaar01} and the hyperluminous X-ray source ($L_{\rm X}>10^{41}~\rm ergs\,s^{-1}$) M82 X-1 in starburst galaxy M82 \citep{pash14}. The cool thermal emission components found in the X-ray spectral analysis of some ULXs seem to be well interpreted by the IMBH accretion disk models \citep{mill04}. The detection of transient radio emission from the candidate IMBH ESO 243-49 HLX-1 constrained the mass of the BH to be $9\times10^3-9\times10^4~M_\odot$ \citep{webb12}, which is in the expected mass range of the IMBHs.

To produce a stable strong X-ray luminosity of an ULX, the IMBH should accrete material from a tidally captured/exchange encountered main sequence (MS) star via a Roche lobe overflow (RLOF). \cite{hopm04} argued that an IMBH in the center of MGG-11 can capture a passing MS star or giant in a tidal event and produce a ULX source-like luminosity via a RLOF of the donor star after orbital circularization in a timescale longer than $10^7~\rm yr$. Detailed stellar evolution simulations indicated that the accretion of an IMBH from its companion star could power an X-ray luminosity exceeding $10^{40}~\rm ergs\,s^{-1}$ in a timescale of several Myr \citep{port04,li04}. Compared with those stellar-mass BHs, the higher BH masses and wider orbits in IMBH X-ray binaries tend to result in a larger accretion disk and a lower temperature of the disk, making them most likely to appear as transient sources due to the thermal-viscous instability \citep{kalo04}. Based on 30,000 binary evolution models of IMBH X-ray binaries, \cite{madh06} found that only those IMBH X-ray binaries with a donor-star mass greater than $8~M_\odot$ and initial orbital separations of about $6-30$ times the radii of donor stars can appear as active ULXs with luminosities greater than $10^{40}\rm ~ erg\,s^{-1}$. Employing N-body simulations of young clusters such as MGG-11 of M82, \cite{baum06} reported that IMBHs have a high probability of capturing stars through tidal energy dissipation and then evolving into the IMBH X-ray binaries mentioned above.

At cosmological distances, a broad population of IMBHs may be revealed by multiband observations via space-borne and ground-based gravitational wave (GW) detectors \citep{Jani2020}. In observation, IMBH X-ray binaries can appear as not only ULXs but also low-frequency GW sources detected by space-borne GW detectors such as LISA \citep{amar23}, TianQin \citep{luo16}, and Taiji \citep{ruan20}. If the initial orbital periods are shorter than the bifurcation periods, IMBH binaries consisting of an IMBH and a MS companion star could evolve toward IMBH X-ray binaries with a compact orbit, which may be detected by space-borne GW detectors \citep{port04b,chen20a}. However, it is not optimistic to detect compact IMBH X-ray binaries by LISA, whose maximum number in the Galaxy should be less than ten in a mission duration of 4 yr \citep{chen20a}. Actually, the formation of compact IMBH X-ray binaries strongly depends on the loss mechanism of angular momentum. If an efficient loss mechanism of angular momentum exists, the number of compact IMBH X-ray binaries detected as low-frequency GW sources should significantly increase. 

The adiabatic growth of a supermassive BH at the center of a galaxy by standard accretion of dust and gases could alter the distribution of dark matter (DM) and create a high-density cusp, i.e., the DM spike \citep{gond99,gned04,merr04,sade13}. In fact, rotation significantly increases the density of the DM close to the BH, resulting in the creation of mini-spikes around a spinning IMBH \citep{ferr17}.  When the donor star of an IMBH is orbiting inside a collisionless DM background, the gravitational interaction between the star and the DM will slow down the star, resulting in the effect of dynamical friction (DF). The minispikes would give rise to an efficient DF effect, which then influences the orbital evolution of IMBH binaries. \cite{eda2015} discovered that the DF of DM could produce a significant influence on the GW waveform of extreme mass ratio inspirals.  Compared to binary systems driven purely by GW, DM DF can lead to different evolutionary tracks, relying on local DM density and orbital separation \citep{Gomez2017}. Taking into account the DF of DM, the merger time scale of intermediate or extreme mass ratio inspirals would be reduced by $2-3$ orders of magnitude \citep{Yue2019}. Recently, it was proposed that the abnormal fast orbital decay of two low-mass BH X-ray binaries A0620-00 and XTE J1118 + 480 was caused by the DF of DM with special spike indices $\gamma$, providing indirect evidence for DM spikes around stellar-mass BHs \citep{chan2023}. These works imply that the DF of DM would drive the BH X-ray binaries to evolve toward binary systems with compact orbits, playing an important role in influencing their orbital evolution.

Until now a detailed binary evolution model of IMBH binaries including the DF of DM has been missing in the literature. In this work, we investigate the detectability of IMBH binaries as low-frequency GW sources when the DF of DM is included. We describe the stellar evolution code and input physics in Section 2. The detailed simulated results are shown in Section 3. Finally, we give a discussion and conclusion in Sections 4, and 5, respectively.

\section{Stellar evolution code and input physics}
\subsection{Stellar evolution code}
To diagnose whether some IMBH binaries can evolve toward low-frequency GW sources, we utilize a binary update version in Modules for Experiments in Stellar Astrophysics code \citep[MESA, r12115,][]{paxt11,paxt13,paxt15,paxt18,paxt19} to calculate the evolution of IMBH binaries consisting of an IMBH and a MS companion star. The BH is assumed to be a point mass with a mass of $M_{\rm BH}=1000~M_{\odot}$, and the companion star is considered to be a population I star with a solar composition, in which $X=0.70, Y=0.28, Z=0.02$. The MESA code can simulate the nuclear synthesis of the MS star and the orbital evolution of the system. Tidal interactions that act on an MS star near RLOF can circularize the orbit on a time scale of $\sim 10^{4}~\rm yr$ \citep{clar97}, which is much shorter than the evolutionary timescale before RLOF. Furthermore, mass transfer will also circularize the orbit with residual eccentricity on a time scale much shorter than the mass transfer timescale \citep{sepi10}. Therefore, the orbits of all IMBH binaries are thought to be circular during their evolution.

If the donor star orbits within the tidal radius of the IMBH, the star may be destroyed as a tidal disruption event (TDE). The tidal radius of an IMBH is \citep{hopm04}
\begin{equation}
R_{\rm{t}}=\left(\frac{M_{\rm{BH}}}{M_{\rm{d}}}\right)^{1/3}R_{\rm d},
\end{equation}
where $M_{\rm d}$ and $R_{\rm d}$ are the mass and radius of the donor star, respectively. Once the donor star spirals in the BH's tidal radius, we stop the calculation because of the completion of the X-ray binaries phase. Otherwise, the detailed stellar evolution model by the MESA code continues running until the time step is less than a minimum time step limit or the stellar age exceeds the Hubble time (14 Gyr).

Once nuclear evolution initiates an extra C/O burning during and after He burning, the Type 2 opacities are adopted. We take a mixing length parameter as $ \alpha = l/H_{\rm p} = 2$, where $l$ and $H_{\rm p}$ are the mixing length and the height of the local pressure scale of the companion star of the MS, respectively. In the numerical calculation, the time step options with $mesh_{-}delta_{-}coeff=1.0$, and $varcontrol_{-}target=10^{-4}$ are used. 

During mass transfer, the mass growth rate of the IMBH is limited by the Eddington accretion rate. The excess materials are assumed to be ejected in the form of isotropic winds in the vicinity of the IMBH, carrying away its specific orbital-angular momentum. Actually, our calculations indicate that a super-Eddington mass transfer is impossible for a low-mass or intermediate-mass donor star. Furthermore, we also consider three other mechanisms of angular momentum loss, including gravitational radiation (GR), magnetic braking, and the DF of DM (see also Section 2.2). If the donor star develops a convective envelope and has a radiative core, the standard magnetic braking model with $\gamma_{\rm mb}=3$ given by \cite{rapp83} is adopted.

\subsection{DF of DM}
Similar to supermassive BHs, the density distribution of DM around an IMBH is described as a piecewise function as follows \citep{sade13,Lacroix2018} 
\begin{equation}
	\rho_{\rm DM}=
	\begin{cases}
		0,&r\leq 2R_{\rm sh},\\
		\rho_{0}(\dfrac{r}{r_{\rm sp}})^{-\gamma},&2R_{\rm sh}<r\leq r_{\rm sp},\\
		\rho_{0},& r > r_{\rm sp},
   \end{cases}
   \label{DM2}
\end{equation}
where $r$ is the radial distance from the BH, $R_{\rm sh}=2GM_{\rm BH}/c^{2}$ ($G$ is the gravitational constant and $c$ is the speed of light in vacuo) is the Schwarzschild radius, $r_{\rm sp}$ is the spike radius of DM. Beyond the spike radius, the DM density around the BH follows a constant distribution with a density of $\rho_{0}$, which may have a tiny difference in different positions of the Galaxy. However, this discrepancy is trivial in those regions apart from the Galactic center \citep{mcmi17}. In the calculation, the density of the DM background is taken to be $\rho_{0}=\rho_{\odot}$ \citep{Lacroix2018}, where $\rho_{\odot}=0.33 \pm 0.03~\rm GeV\,cm^{-3}$ is the DM density in the vicinity of Sun. We employ the standard assumption $r_{\rm sp}=0.2r_{\rm in}$ \citep{Fields2014,eda2015}, where $r_{\rm in}$ is the influence radius of a BH. The total mass of DM inside the influence radius is $2M_{\rm BH}$ \citep{merr04}, that is 
\begin{equation}
2M_{\rm BH}=\int_{0}^{r_{\rm in}} 4\pi r^{2} \rho_{\rm DM} {\rm d}r.
\label{rin}
\end{equation}
We plot the DM density profiles around an IMBH of $1000~M_\odot$ in Figure 1. It is clear that the minispike around an IMBH leads to a distribution of high DM density, in which a higher spike index tends to produce a higher DM density in the vicinity of the IMBH. For a system consisting of a $1000~M_\odot$ IMBH and a $1~M_\odot$ MS star in an orbit of 1.0 day, the DM densities with $\gamma=1.0-2.0$ at the position of the donor star are in the range of $\sim10^6-10^{13}~\rm GeV\,cm^{-3}$, which are $6-13$ orders of magnitude higher than $\rho_0$.

\begin{figure}
\centering
\includegraphics[width=1.1\linewidth,trim={0 0 0 0},clip]{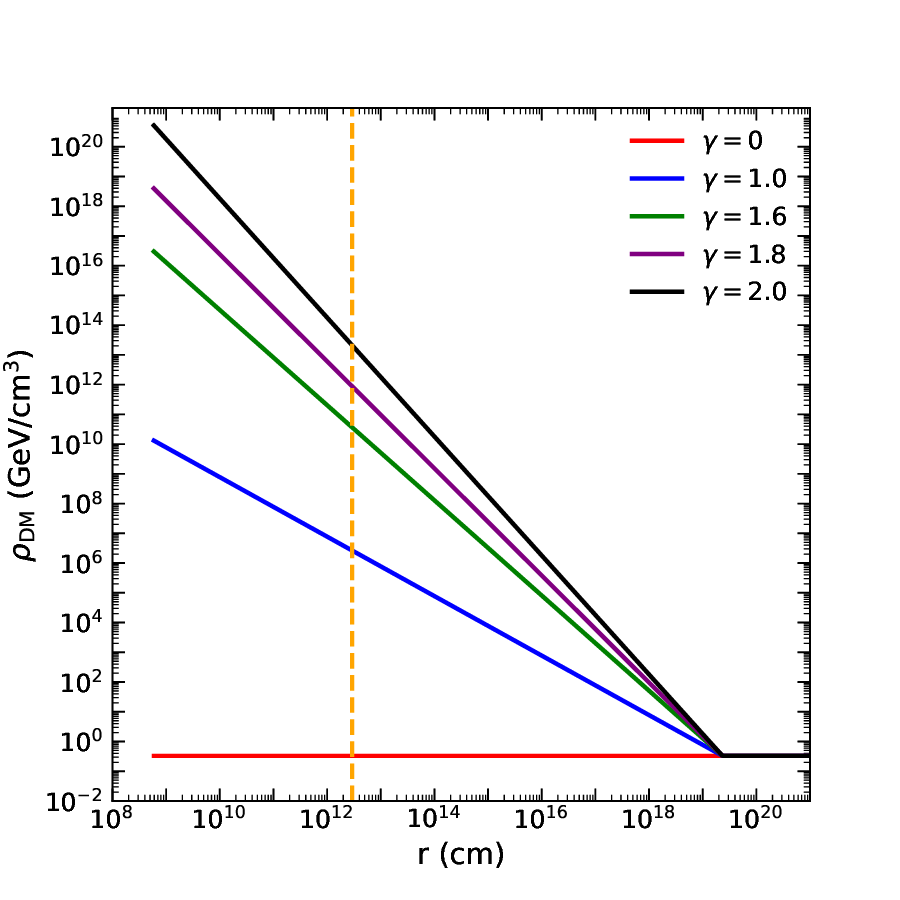}
\caption{The DM density profiles with different spike indices around an IMBH of $M_{\rm BH}=1000~ M_{\odot}$. The vertical orange dashed line represents a radius of $r=2.93\times10^{12}~\rm cm$, which is the orbital separation of a system consisting of a $1000~M_\odot$ IMBH and a $1~M_\odot$ MS star in an orbit of 1.0 day.}\label{fig:DMdensity}
\end{figure}

The orbital energy dissipation rate due to the DF of DM is given by \citep{Yue2019}
\begin{equation}
\dot{E}_{\rm DF}=-\frac{4\pi G^{2} \mu^{2} \rho_{\rm DM} \xi(\sigma) \ln\Lambda}{v},
\end{equation}
where $\mu=M_{\rm BH}M_{\rm d}/(M_{\rm BH}+M_{\rm d})$ is the reduced mass of the system, $\xi(\sigma)$ is a numerical factor determining by the distribution function and the velocity
dispersion of DM, $\ln\Lambda \approx \ln\sqrt{M_{\rm BH}/M_{\rm d}}$ is called the Coulomb logarithm \citep{Kavanagh2020}, $v$ is the orbital velocity of the companion star.

According to $\dot{J}=\dot{E}/\Omega$ ($\Omega$ is the orbital angular velocity), the loss rate of orbital angular momentum due to the DF of DM can be written as \citep{qin24}
\begin{equation}
\dot{J}_{\rm DF}=-\frac{G^{2}\mu^{2}(1+q)\rho_{\rm DM}\xi(\sigma){\rm ln}\Lambda P_{\rm orb}^{2}}{\pi a},
\label{eq:jdot5}
\end{equation}
where $q=M_{\rm d}/M_{\rm BH}$, $P_{\rm orb}$, and $a$ are the mass ratio, orbital period, and orbital separation of IMBH binaries, respectively.

\subsection{Detectability of IMBH binaries as low-frequency GW sources}
Once those IMBH binaries evolve into systems with a narrow orbit, they will produce low-frequency GW signals with a frequency of $f_{\rm gw}=2/P_{\rm orb}$, which may be detected by spaceborne GW detectors such as LISA. The maximum GW frequency derivative is $f_{\rm gw,max}\sim 10^{-14}~\rm Hz\,s^{-1}$ (see also Figure 9. In fact, $f_{\rm gw,max}\sim 10^{-18}~\rm Hz\,s^{-1}$ for those semi-detached systems), hence the frequency change is $\Delta f_{\rm gw}\sim10^{-6}~\rm Hz$ in a LISA mission duration of $T=4~\rm yr$. Therefore, these low-frequency GW sources can be thought of in a circular orbit with monotonically varying frequency. Considering an average orbital orientation and polarization of a binary GW source, the GW amplitude at a distance of $d$ is \citep{evan87}
\begin{equation}
h_0=\sqrt{\frac{32\pi^{4/3}G^{10/3}}{5}}\frac{f_{\rm gw}^{2/3}\mathcal{M}^{5/3}}{c^4d},    
\end{equation}
where $\mathcal{M}=(M_{\rm BH}M_{\rm d})^{3/5}/(M_{\rm BH}+M_{\rm d})^{1/5}$ is the chirp mass. By integrating the power in the signal over a mission duration, the characteristic strain can be derived as $h_{\rm c}=\sqrt{2f_{\rm gw}T}h_0$ \citep{finn00,taur18}. Therefore, the characteristic strain of low-frequency GW signals produced by IMBH binaries is given by \citep{chen20a}
\begin{equation}
	\begin{aligned}   
		h_{c} \approx 3.75 \times 10^{-20} \left (\frac{f_{\rm gw}}{1 ~\rm mHz} \right)^{7/6} 
		\left (\frac{\mathcal{M}}{1~M_{\odot}}\right )^{5/3} \left (\frac{10~\rm kpc}{d}\right).
	\end{aligned}
    \label{equation:hc6}
\end{equation}
For simplicity, those IMBH binaries are determined to be low-frequency GW sources that can be detected by LISA when the calculated characteristic strain exceeds the LISA sensitivity curve given by an analytic estimation \citep{robs19}. We emphasize that equation (7) is not the characteristic strain in the frequency domain. However, it should be noted that our calculated evolutionary track in the characteristic strain vs. instantaneous GW frequency diagram could be representative of the order of magnitude of the expected strain during the evolution of the IMBH binaries.

\section{Simulated results}
It strongly depends on the loss rate of angular momentum whether or not IMBH-MS binaries can evolve toward low-frequency GW sources. According to Equation (\ref{eq:jdot5}), the loss rate of angular momentum is closely related to three parameters, including $M_{\rm d}$, $\rho_{\rm DM}$ and $P_{\rm orb}$. Therefore, we first investigate the influence of these three parameters on the evolution of IMBH-MS binaries.

\subsection{Influence of donor-star masses}
\begin{figure}
\centering
\includegraphics[width=1.15\linewidth,trim={0 0 0 0},clip]{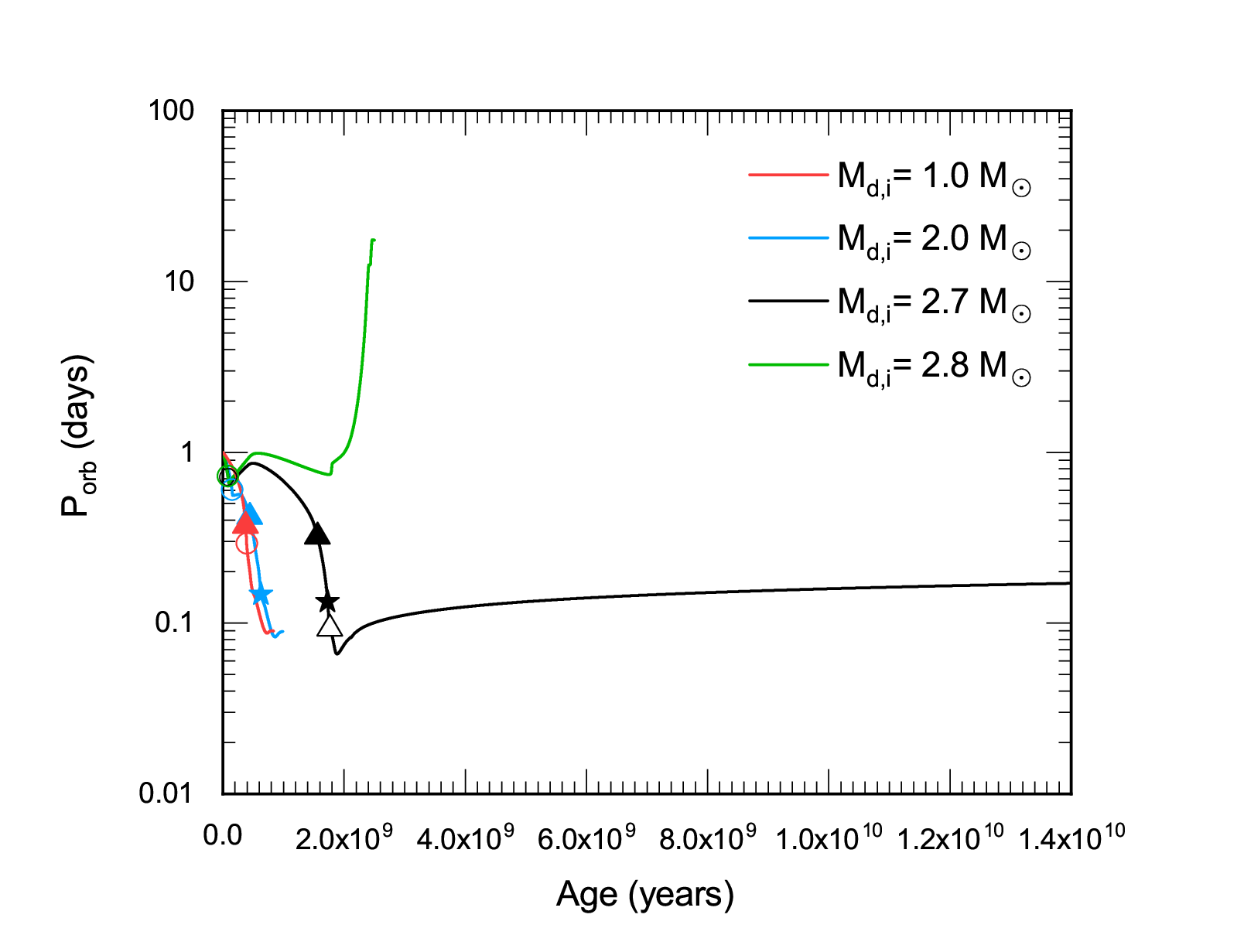}
\includegraphics[width=1.15\linewidth,trim={0 0 0 0},clip]{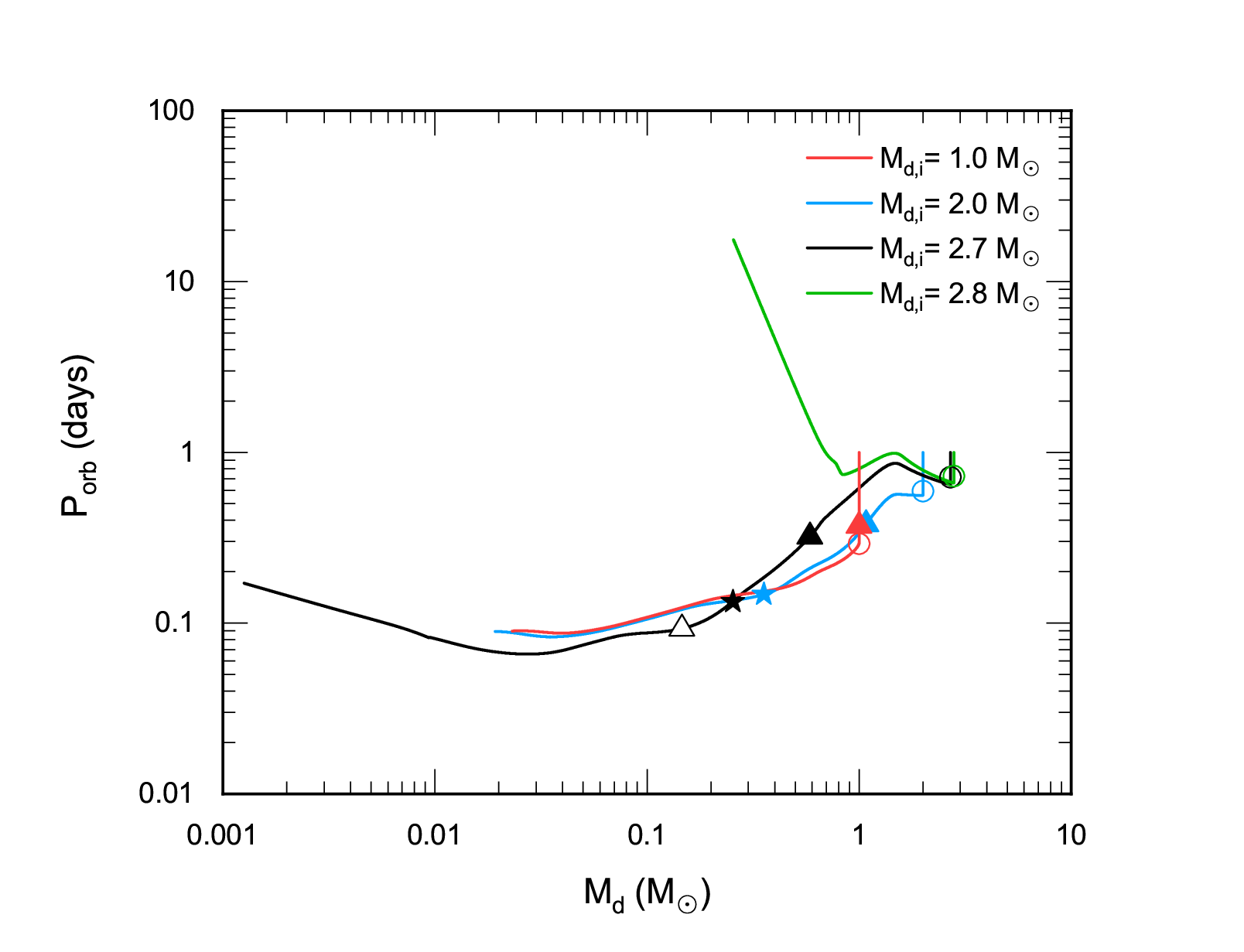}
\caption{Evolution of IMBH-MS binaries with an initial orbital period of $P_{\rm orb,i}=1.0~\rm day$ and different initial donor-star masses in the orbital period vs. stellar age diagram (top panel) and orbital period vs. donor-star mass diagram (bottom panel) when $\gamma=1.60$. The open circles, solid triangles, and solid stars indicate the beginning of RLOF, the onset when IMBH binaries are visible by LISA at distances of 1 kpc and 10 kpc, respectively. These sources are invisible for LISA at a distance of 1 kpc after those open triangles.} \label{fig:1.60a}
\end{figure}

Figure \ref{fig:1.60a} shows the evolution of four IMBH-MS binaries with $M_{\rm d,i}=1.0-2.8~M_{\odot}$, $P_{\rm orb,i}=1~\rm day$, and $\gamma=1.6$. For a relatively small spike index of 1.6, those systems with $M_{\rm d,i}\ge 2.8~M_{\odot}$ eventually evolve into wide-orbit systems. In contrast, those IMBH binaries with a low donor star mass ($M_{\rm d,i}\le2.7~M_{\odot}$) can evolve into compact X-ray binaries, which are visible for LISA as low-frequency GW sources in a timescale of $400-700$ Myr. When these sources are visible for LISA at a distance of 1 kpc, the donor-star masses decrease to $M_{\rm d}\sim0.7-1.0~M_{\odot}$, and the orbital periods decrease to $\la0.4~\rm days$, corresponding to a GW frequency of $\ga0.06~\rm mHz$. 

The total orbital angular momentum of an IMBH-MS binary is $J=M_{\rm BH}M_{\rm d}/(M_{\rm BH}+M_{\rm d})a^2\Omega$. Ignoring the mass loss during the mass transfer, i.e., $\dot{M}_{\rm BH}=0$, from Kepler's third law ($G(M_{\rm BH}+M_{\rm d})/a^3 = 4\pi^2/P_{\rm orb}^2$) we can derive 
\begin{equation}
    \frac{\dot{P}_{\rm orb}}{P_{\rm orb}}=3\frac{\dot{J}}{J}-3\frac{\dot{M_{\rm d}}}{M_{\rm d}}(1-q),
    \label{equation:pdot7}
\end{equation}
where $q=M_{\rm d}/M_{\rm BH}$ is the mass ratio of the binary. Since $\dot{M}_{\rm d}<0$ and $q<1$, the second term on the right-hand side of equation (\ref{equation:pdot7}) is positive. Therefore, mass transfer causes an orbital expansion effect when the mass is transferred from the less massive donor star to the more massive BH. However, the loss of angular momentum would lead to an orbital decay effect (see also the first term on the right-hand side of the equation \ref{equation:pdot7}). As a consequence, the evolutionary fate of an IMBH binary strongly depends on the competition between two effects caused by the mass transfer and the angular momentum loss. A smaller spike index directly leads to a lower DM density, then indirectly produces a lower rate of angular momentum loss due to the DF of DM. Therefore, a small spike index cannot drive those IMBH binaries with massive donor stars to evolve toward low-frequency GW sources. In Figure \ref{fig:1.60a}, the early evolution of three systems with $M_{\rm d,i}=1.0,2.0$, and $ 2.7~M_{\odot}$ is dominated by the DF of DM. Hence they experience a rapid orbital shrinkage and evolve into compact orbit binaries. However, the orbital expansion effect in the system with $M_{\rm d,i}=2.8~M_\odot$ gradually conquers the orbital decay effect caused by the DF of DM, driving it to evolve toward a wide orbit system.

\begin{figure}
\centering
\includegraphics[width=1.15\linewidth,trim={0 0 0 0},clip]{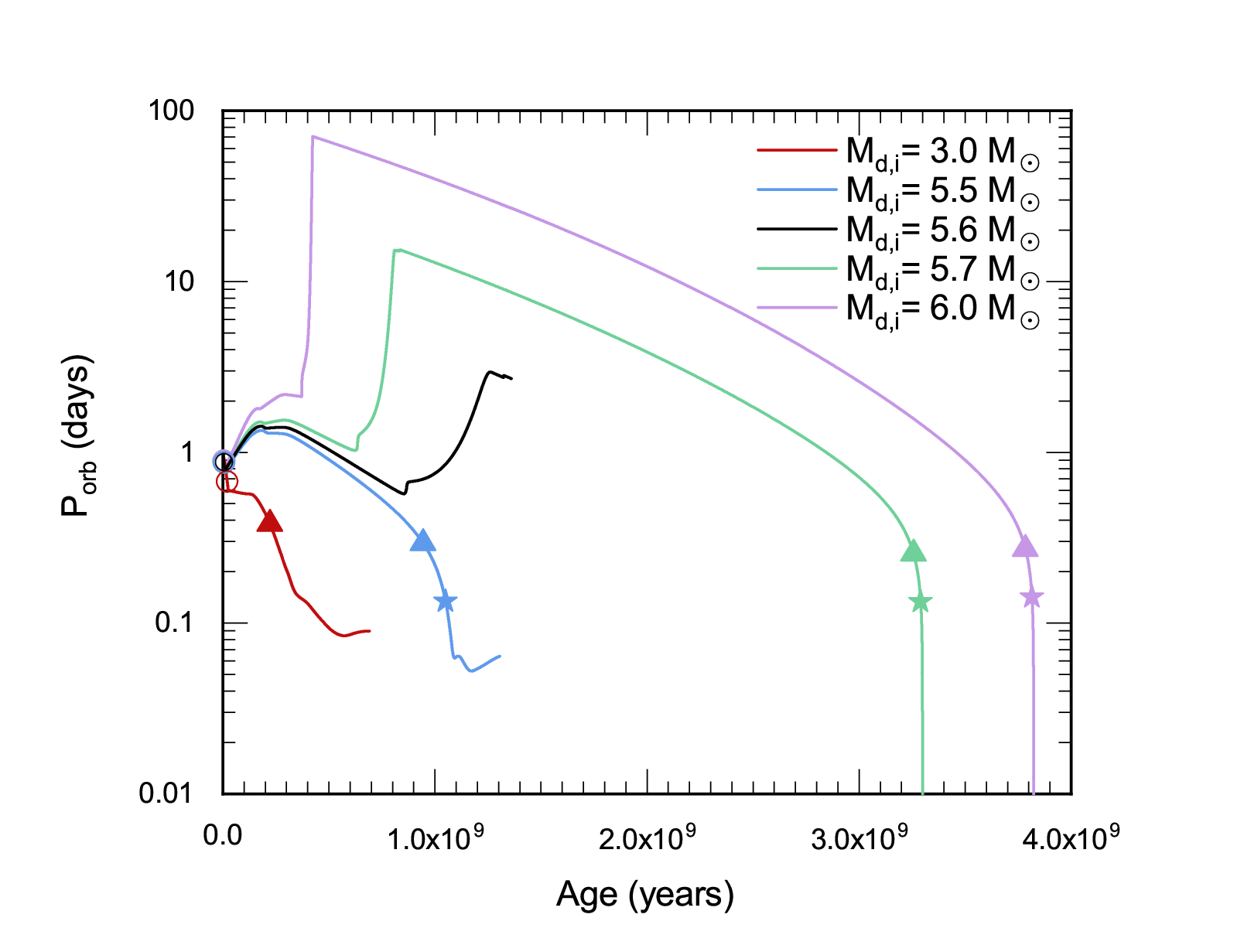}
\includegraphics[width=1.15\linewidth,trim={0 0 0 0},clip]{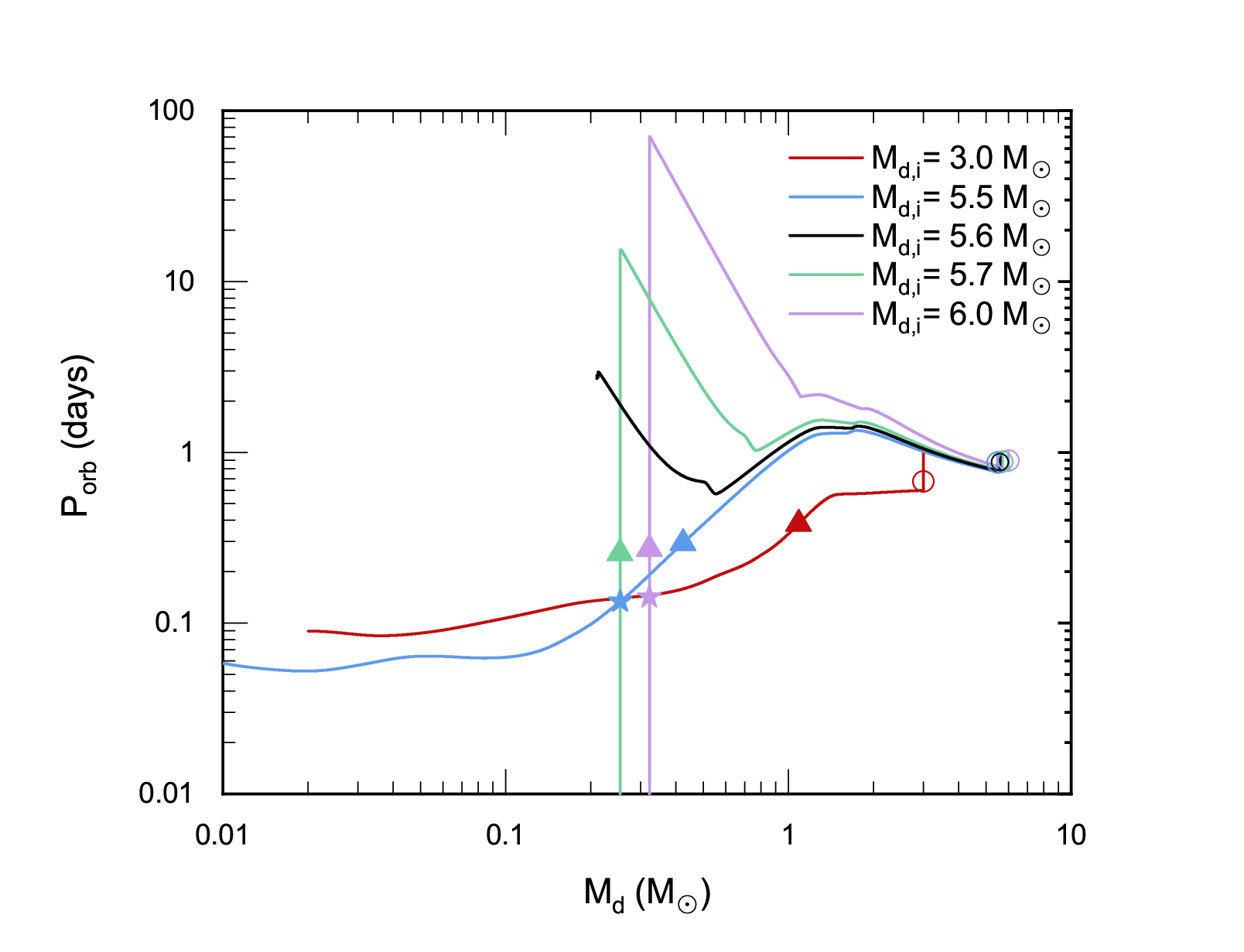}
\caption{Evolution of IMBH-MS binaries with $P_{\rm orb, i}=1.0~\rm day $ and different massive donor stars in the orbital period vs. stellar age diagram (top panel) and orbital period vs. donor-star mass diagram (bottom panel) when $\gamma=1.80$. The meanings of those symbols are the same as Figure \ref{fig:1.60a}. } \label{fig:1.80a}
\end{figure}

In Figure \ref{fig:1.80a}, we adopt a large spike index ($\gamma=1.8$) to investigate the evolution of the IMBH-MS binaries with $M_{\rm d,i}=3.0-6.0~M_{\odot}$ and $P_{\rm orb, i}=1~\rm day$. Their evolutionary tendency is very similar to that in Figure \ref{fig:1.60a}, in which a lower initial donor-star mass tends to result in the formation of more compact X-ray binaries and low-frequency GW sources. As in $\gamma=1.80$, a higher DM density produces a higher rate of angular momentum loss, and hence, the DF of DM can dominate the evolution of some IMBH binaries with massive donor stars. Therefore, two systems with $M_{\rm d,i}=5.5$ and $3.0~M_{\odot}$ can evolve toward compact IMBH X-ray binaries and low-frequency GW sources. When those IMBH X-ray binaries appear as low-frequency GW sources at a distance of 1 kpc, the donor-star masses are distributed in a range of $\sim0.4-1.0~M_{\odot}$ and the orbital periods are $\la0.4~\rm days$. Three IMBH binaries with relatively high donor-star mass ($M_{\rm d,i}=5.6,5.7$, and $6.0~M_{\odot}$) first evolve into systems with wide orbits because of the strong orbital expansion effect caused by mass transfer. However, the two systems with $M_{\rm d,i}=5.7$ and $6.0~M_{\odot}$ would decouple their Roche lobes and become detached systems consisting of an IMBH and a white dwarf (WD) within a Hubble timescale. Subsequently, GR drives these two systems to experience a rapid orbital shrinkage and eventually evolve into low-frequency GW sources that are visible for LISA at a distance of 1 kpc and 10 kpc in sequence.

\begin{figure}
\centering
\includegraphics[width=1.15\linewidth,trim={0 0 0 0},clip]{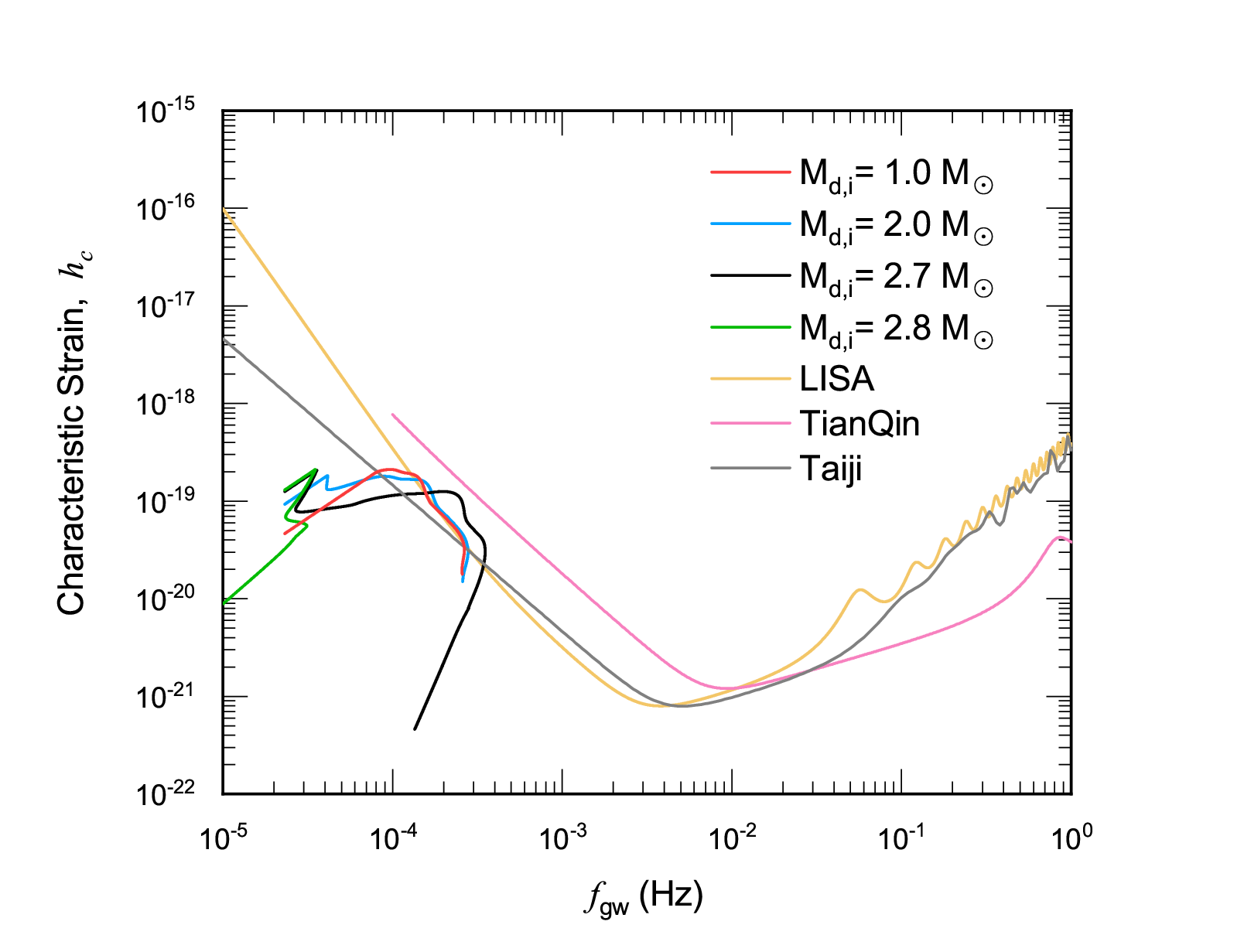}
\includegraphics[width=1.15\linewidth,trim={0 0 0 0},clip]{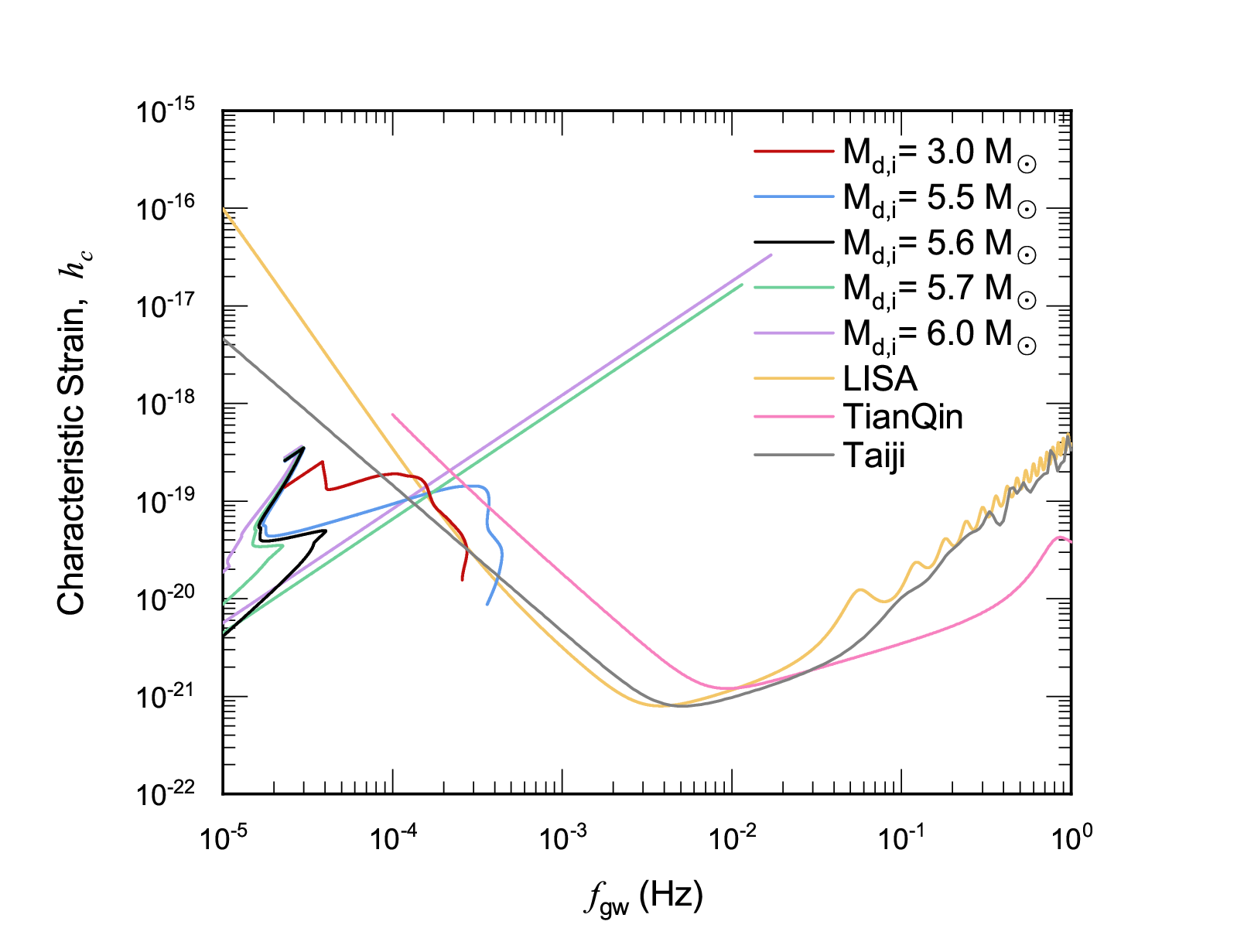}
\caption{Evolution of IMBH binaries with $P_{\rm orb,i}=1~\rm day$ and different donor-star masses in the characteristic strain vs. GW frequency diagram when $\gamma=1.60$ (top panel) and $\gamma=1.80$ (bottom  panel). The blue curve is the sensitivity curve of LISA originating from the numerical calculation of a mission duration of 4 yr. The red and green curves are the sensitivity curves of TianQin \citep{Wang2019} and Taiji \citep{ruan20}, respectively.} \label{fig:LISAa}
\end{figure}

To test whether those IMBH-MS binaries can evolve toward low-frequency GW sources, Figure \ref{fig:LISAa} depicts the evolutionary tracks of the systems in Figures 1 and 2 in the characteristic strain versus GW frequency. In the upper panel, three systems with $M_{\rm d,i}=1.0, 2.0,~{\rm and}~2.7~M_{\odot}$ and $P_{\rm orb,i}=1.0~\rm day$ can evolve to penetrate the sensitivity curves of LISA and Taiji at a GW frequency of $\sim 0.1~\rm mHz$, and become visible LISA and Taiji sources when $d=10~\rm kpc$. Because of a short arm length, it is difficult for TianQin to detect the GW signals from these three systems. In the bottom panel, four systems with massive donor stars can evolve into visible LISA and Taiji sources (TianQin can only detect three systems among these four systems, the system with $M_{\rm d,i}=3.0~ M_{\odot}$ is invisible at a distance of 10 kpc) except for the system with $M_{\rm d,i}=5.6~ M_{\odot}$. In the stage of low-frequency GW sources, two systems with $M_{\rm d,i}=5.7~{\rm and}~6.0~M_{\odot}$ have the same slope as $\Delta({\rm log}h_{\rm c})/\Delta({\rm log}f_{\rm gw})=7/6$. This is because these two systems already evolve into detached IMBH-WD binaries, in which there exists a relation as $h_{\rm c}\propto f_{\rm gw}^{7/6}$ for a constant chirp mass (see also Equation \ref{equation:hc6}). 

According to Equation (26) of \cite{robs19}, the signal-to-noise ratio of typical galactic binaries can be written as ${\rm SNR}\approx h_{\rm c}/\sqrt{S_{\rm n}(f)f_{\rm gw}}$ ($S_{\rm n}(f)$ is the noise spectral density of LISA). We calculate the SNR of seven detectable systems in Figure 4. In the upper panel, three systems with $M_{\rm d,i}=1.0, 2.0,~{\rm and}~2.7~M_{\odot}$ have maximum SNRs of $\sim1$, $\sim1$, and $\sim2$ when $d=10~\rm kpc$, respectively. If the detected distance decreases to $1~\rm kpc$, the maximum SNRs of these three systems are $\sim9$, $\sim10$, and $\sim20$, respectively. Therefore, these three systems are difficult for LISA to detect at a long distance of 10 kpc, but can be detected at a close distance of 1 kpc. In the bottom panel, the maximum SNRs of four systems with $M_{\rm d,i}=3.0, 5.5, 5.6~{\rm and}~5.7~M_{\odot}$ are $\sim1$ (this system cannot be detected by LISA), $\sim5$, $\sim1.2\times10^4$, and $\sim1.9\times10^4$ when $d=10~\rm kpc$, respectively.

\subsection{Influence of spike indices}
\begin{figure}
\centering
\includegraphics[width=1.15\linewidth,trim={0 0 0 0},clip]{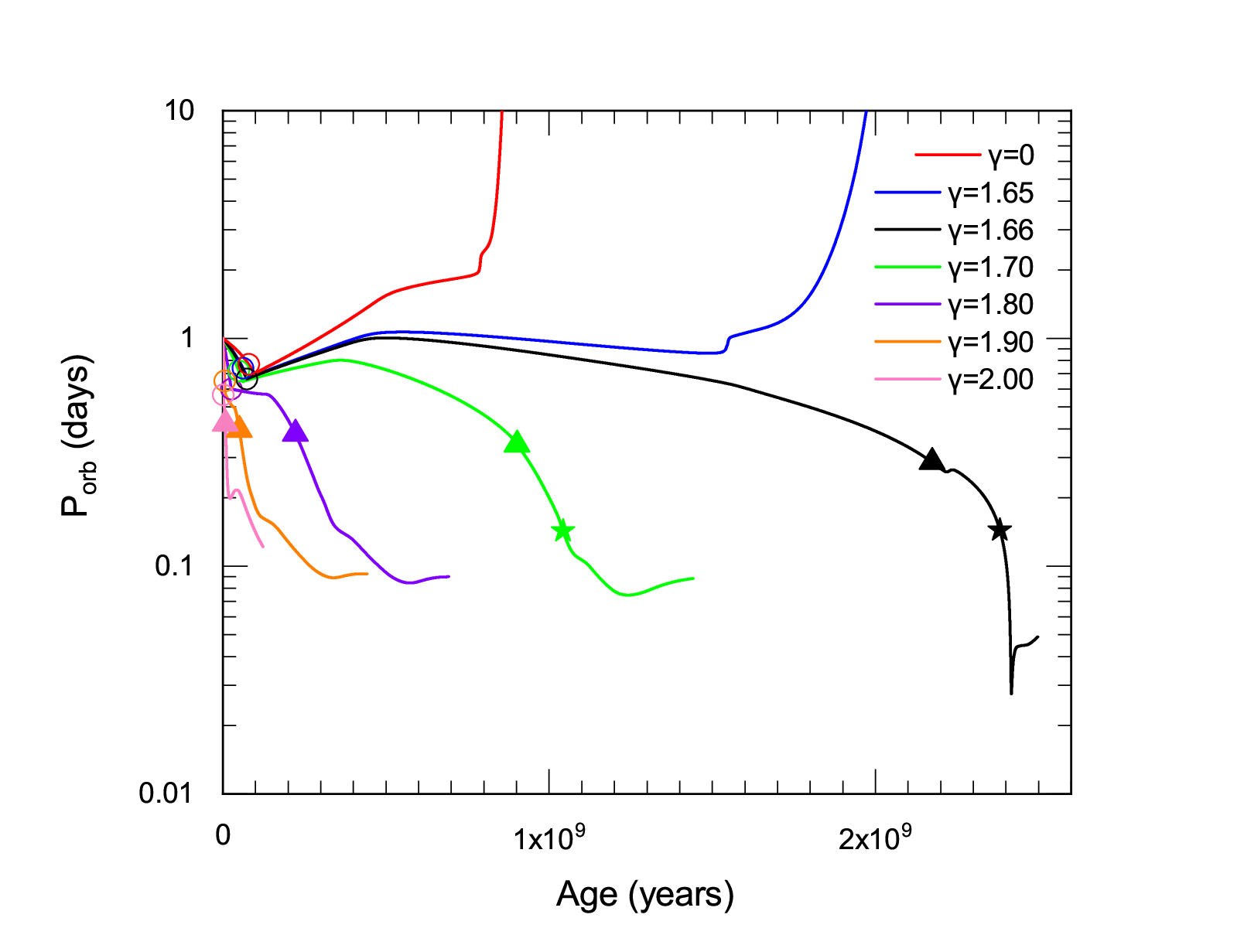}
\caption{Evolution of IMBH-MS binaries with $M_{\rm d,i}=3.0~ M_{\odot}$, $ P_{\rm orb,i}=1.0 ~\rm day$, and different spike indices in the orbital period vs. stellar age diagram. The meanings of those symbols are same to Figure \ref{fig:1.60a}.} \label{fig:indices}
\end{figure}

According to equations (\ref{DM2}) and (\ref{eq:jdot5}), a higher spike index tends to produce a higher density of DM and a higher loss rate of angular momentum due to DF of DM. Therefore, the spike index plays an important role in influencing the evolutionary fates of the IMBH-MS binaries. Taking into account different spike indices, Figure \ref{fig:indices} illustrates the evolution of IMBH-MS binaries with $M_{\rm d,i}=3.0~ M_{\odot}$ and $ P_{\rm orb,i}=1.0 ~\rm day$ in the orbital period versus stellar age diagram. It is clear that there exists a critical spike index $\gamma_{\rm cr} = 1.66$, under which those IMBH-MS binaries cannot evolve toward low-frequency GW sources. This critical spike index is slightly higher than that ($\gamma_{\rm cr} = 1.58$) in stellar-mass BH X-ray binaries \citep{qin24}. Meanwhile, a higher spike index naturally results in a higher orbital period derivative (see also the slopes of the evolutionary curves) and a shorter evolutionary timescale. When five IMBH X-ray binaries are visible for LISA within a distance of 1 kpc, their ages are 2.2, 0.9, 0.25, 0.1, and 0.01 Gyr for $\gamma=1.66,1.7,1.8,1.9$, and 2.0, respectively. Under an assumption of standard magnetic braking (i.e., $\gamma=0$), the IMBH-MS binary with an intermediate-mass donor star cannot evolve into a compact orbit system. 

Among those spike indices higher than $\gamma_{\rm cr}$, a smaller spike index tends to form an IMBH X-ray binary with a shorter minimum orbital period. This is because a smaller spike index leads to a lower DM density at the position of the donor star, inducing a lower loss rate of angular momentum due to the DF of DM and a longer evolutionary timescale. Consequently, a longer evolutionary timescale makes the donor star develop a more massive He core in its core, subsequently resulting in a more compact donor star and a correspondingly shorter orbital period \citep[see also][]{qin24}.

\subsection{Influence of initial orbital periods}
\begin{figure}
\centering
\includegraphics[width=1.15\linewidth,trim={0 0 0 0},clip]{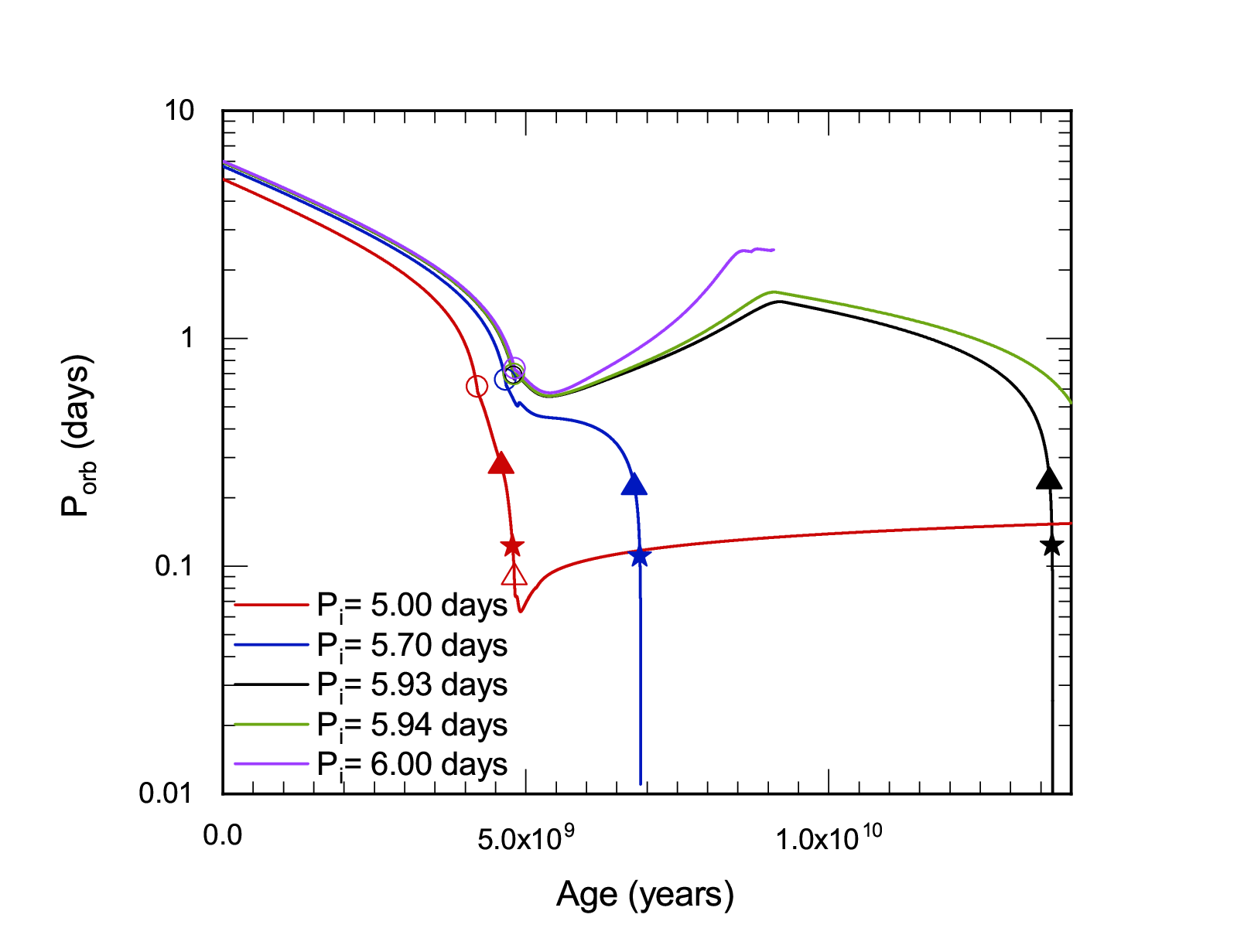}
\includegraphics[width=1.15\linewidth,trim={0 0 0 0},clip]{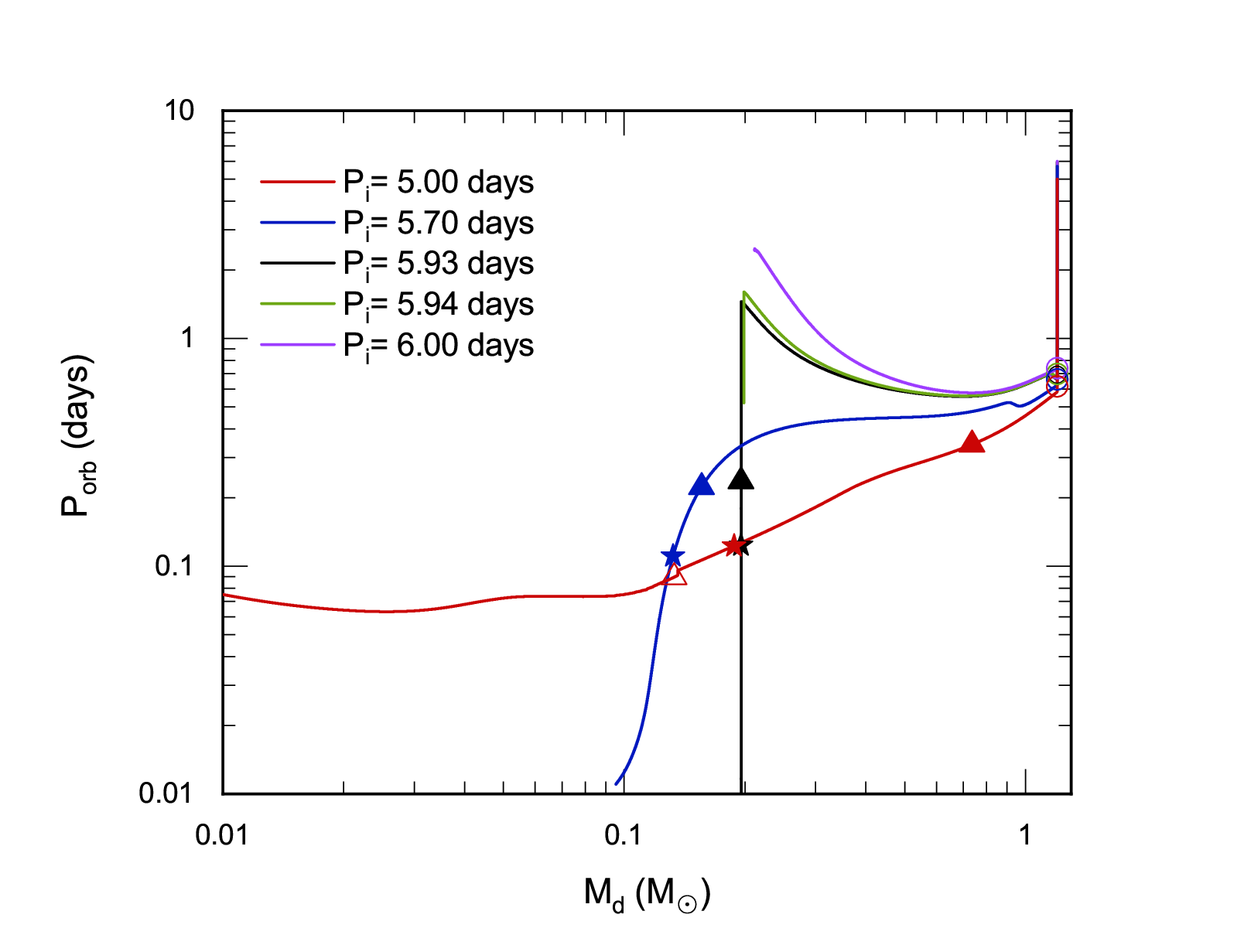}
\caption{ Evolution of IMBH-MS binaries with $M_{\rm d,i}=1.2~ M_{\odot}$, $\gamma=1.60$, and different initial orbital periods in the orbital period vs. stellar age diagram (top panel) and orbital period vs. donor-star mass diagram (bottom panel). The meanings of those symbols are same to Figure \ref{fig:1.60a}.} \label{fig:1.2suna}
\end{figure}

Taking $\gamma=1.60$ and $M_{\rm d,i}=1.2~ M_{\odot}$, Figure \ref{fig:1.2suna} summarizes the evolution of IMBH-MS binaries with different initial orbital periods in the orbital period versus the stellar age diagram and the orbital period versus donor-star mass diagram. There exists a bifurcation period of $P_{\rm orb,i}=5.93~\rm days$, over which those IMBH-MS binaries cannot evolve toward compact orbit systems within a Hubble time. The bifurcation period was defined as the longest initial orbital period in which binary systems consisting of a compact object and a MS donor star can evolve into ultracompact X-ray binaries within a Hubble time \citep{sluy05a,sluy05b}. In principle, the mass transfer from the less massive donor star to the more massive IMBH causes an orbital expansion effect, whereas the angular momentum loss due to the DF of DM results in an orbital decay effect. It depends on the competition between these two effects mentioned above whether IMBH-MS binaries can evolve toward ultracompact orbit systems. As a consequence, the bifurcation period is closely related to the angular momentum loss mechanisms, and a more efficient mechanism of angular momentum loss tends to lead to a longer bifurcation period. When magnetic braking is a dominant mechanism of angular momentum loss, the bifurcation period of an IMBH binary consisting of a $1000~M_\odot$ IMBH and a $1.2~M_\odot$ MS donor star is between 1.9 and 3.2 days according to Figure 5 in \cite{chen20a}, but our simulated bifurcation period is 5.93 days in the same system. This discrepancy originates from different rates for extracting orbital angular momentum, in which the loss rate of angular momentum due to the DF of DM is much higher than that driven by the standard magnetic braking law in \cite{chen20a}.

In Figure \ref{fig:1.2suna}, the IMBH-MS binaries with an initial orbital period shorter than the bifurcation period can evolve toward low-frequency GW sources. When the orbital periods decrease to $\sim0.2~\rm days$ and $\sim0.1~\rm days$, the systems can be detected by LISA at a distance of 1 kpc and 10 kpc, respectively. The low-frequency GW sources evolving from IMBH-MS binaries can be divided into two populations: semidetached systems and detached systems. In the stage of low-frequency GW sources, two compact IMBH X-ray binaries with initial orbital periods of 5.7 and 5.0 days are semidetached systems in which the mass transfer always proceeds, and the systems also appear as X-ray sources. The system with an initial period of 5.93 days first evolves into a detached binary consisting of an IMBH and a pre-WD (with a mass of $\sim0.2~M_\odot$) in a relatively wide orbit with an orbital period of $\sim1~\rm day $. Subsequently, GR drives this detached system to evolve toward low-frequency GW sources after the pre-WD evolves into a WD through a contraction and cooling phase.

\subsection{Parameter space of IMBH-MS binaries forming low-frequency GW sources}
\begin{figure}
\centering
\includegraphics[width=1.15\linewidth,trim={0 0 0 0},clip]{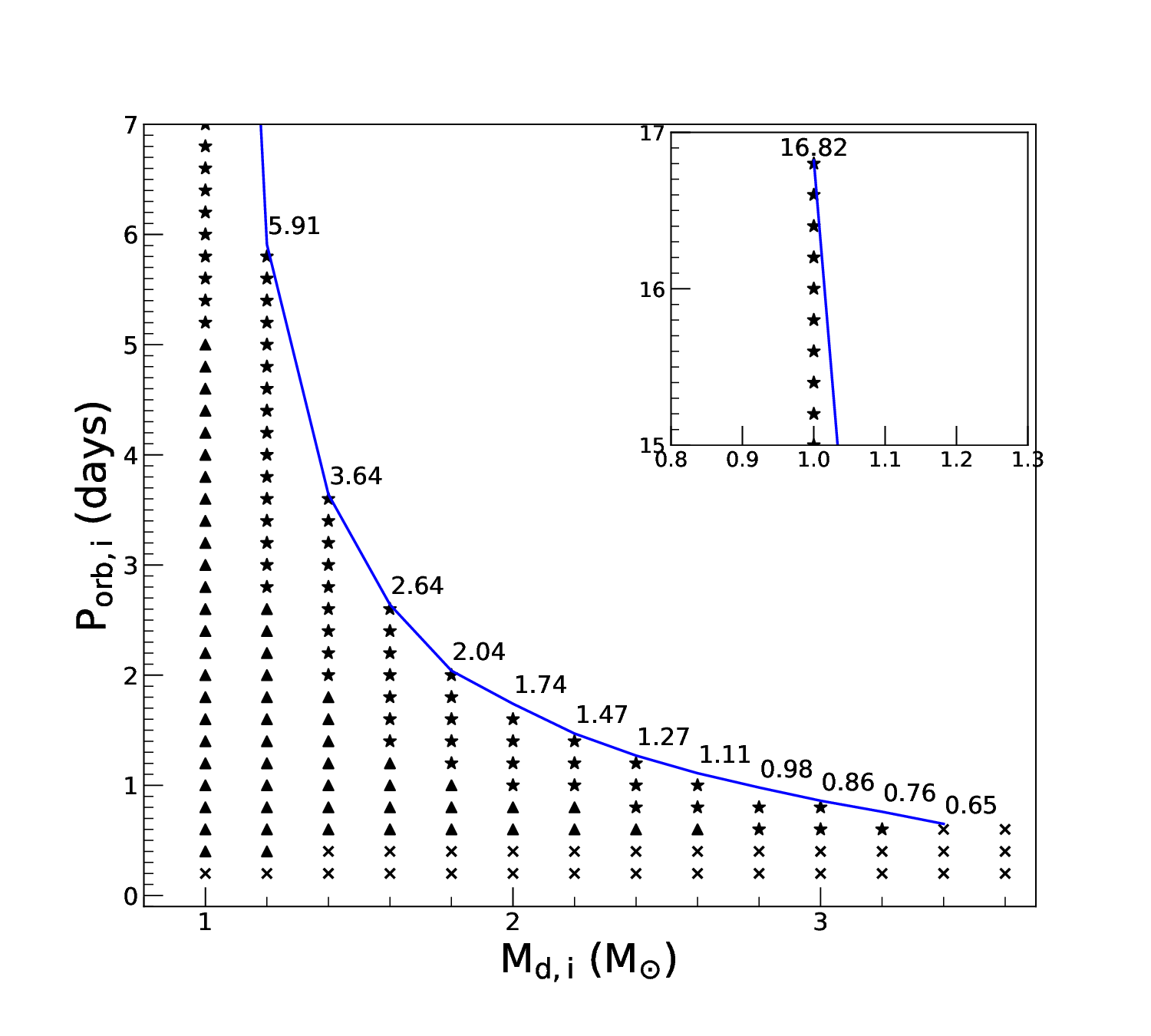}
\caption{Parameter space distribution of IMBH-MS binaries that can evolve toward low-frequency GW sources detected by LISA in the initial orbital period vs. initial donor-star mass diagram. We adopt a constant spike index of DM as $\gamma=1.60$. The solid curve represents the bifurcation periods of IMBH binaries with different donor-star masses.  The solid circles and stars correspond to those IMBH binaries that will evolve toward low-frequency GW sources detected by LISA within a distance of $d =1$ and $10~\rm kpc$, respectively. The crosses stand for those binaries in which the RLOF already occurs at the beginning of binary evolution. When the initial periods are above the bifurcation period, those systems will evolve toward IMBH binaries with long periods.} \label{fig:parameter}
\end{figure}

Using a fixed spike index of $\gamma=1.60$, the evolutionary fates of the IMBH-MS binaries with different initial orbital periods and initial donor-star masses are summarized in Figure \ref{fig:parameter}. Those IMBH binaries with $M_{\rm d,i}=1.0-3.4~ M_{\odot}$ and $P_{\rm orb,i}=0.65-16.82~ \rm days$ are potential visible LISA sources at a distance of $d=10~\rm kpc$. This parameter space is much larger than that given by the standard MB law \citep[see also Figure 5 in][]{chen20a}. The upper boundary of the initial periods corresponds to the bifurcation periods. As the donor-star mass grows, the bifurcation period decreases. This is because a higher donor-star mass produces a higher mass-transfer rate, resulting in a stronger orbital expansion effect. Similarly, those IMBH-MS binaries with higher donor-star mass are difficult to evolve toward compact orbit systems, yielding the right boundary. Those IMBH-MS binaries with an initial orbital period shorter than the bottom boundary cannot evolve toward low-frequency GW sources within a Hubble time or already fill their Roche lobe at the beginning of binary evolution. The left boundary originates from the law that a lower donor-star mass requires a longer evolutionary timescale, and hence, those systems cannot evolve into compact orbit systems within a Hubble timescale. When $M_{\rm d,i}=1.0$, 2.0, and $3.0~ M_\odot$, the bifurcation periods are 16.82, 1.74, and 0.86 days, respectively. Our bifurcation periods are slightly longer than those in \cite{chen20a}, which are 3.2, 1.4, and $0.7~\rm days$, respectively. Similarly to the previous subsection, this implies that the loss rate of angular momentum given by the DF of DM is higher than that provided by the standard MB law.

\section{Discussion}

\subsection{Evolution of companion stars in HR diagram}
\begin{figure}
\centering
\includegraphics[width=1.15\linewidth,trim={0 0 0 0},clip]{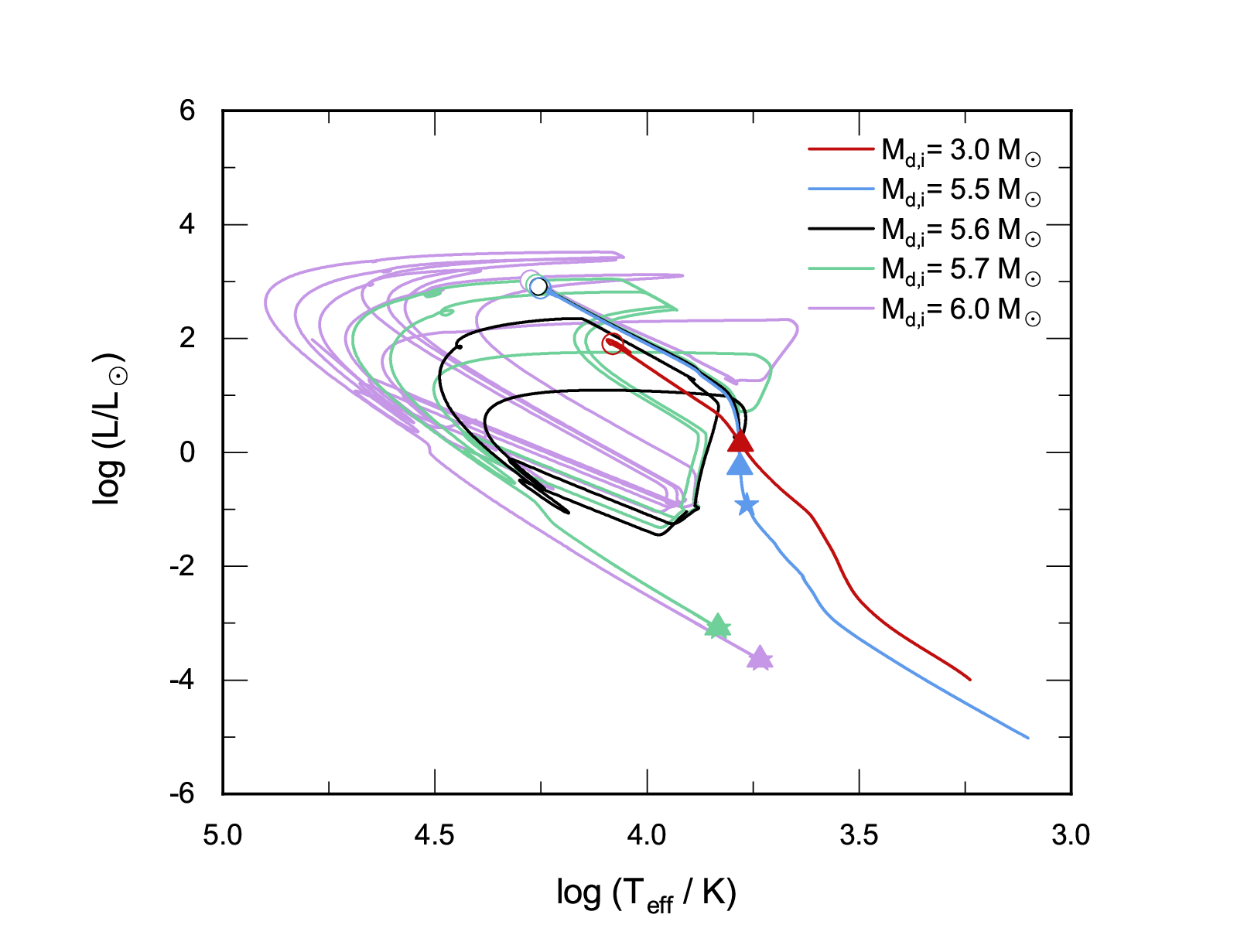}
\includegraphics[width=1.15\linewidth,trim={0 0 0 0},clip]{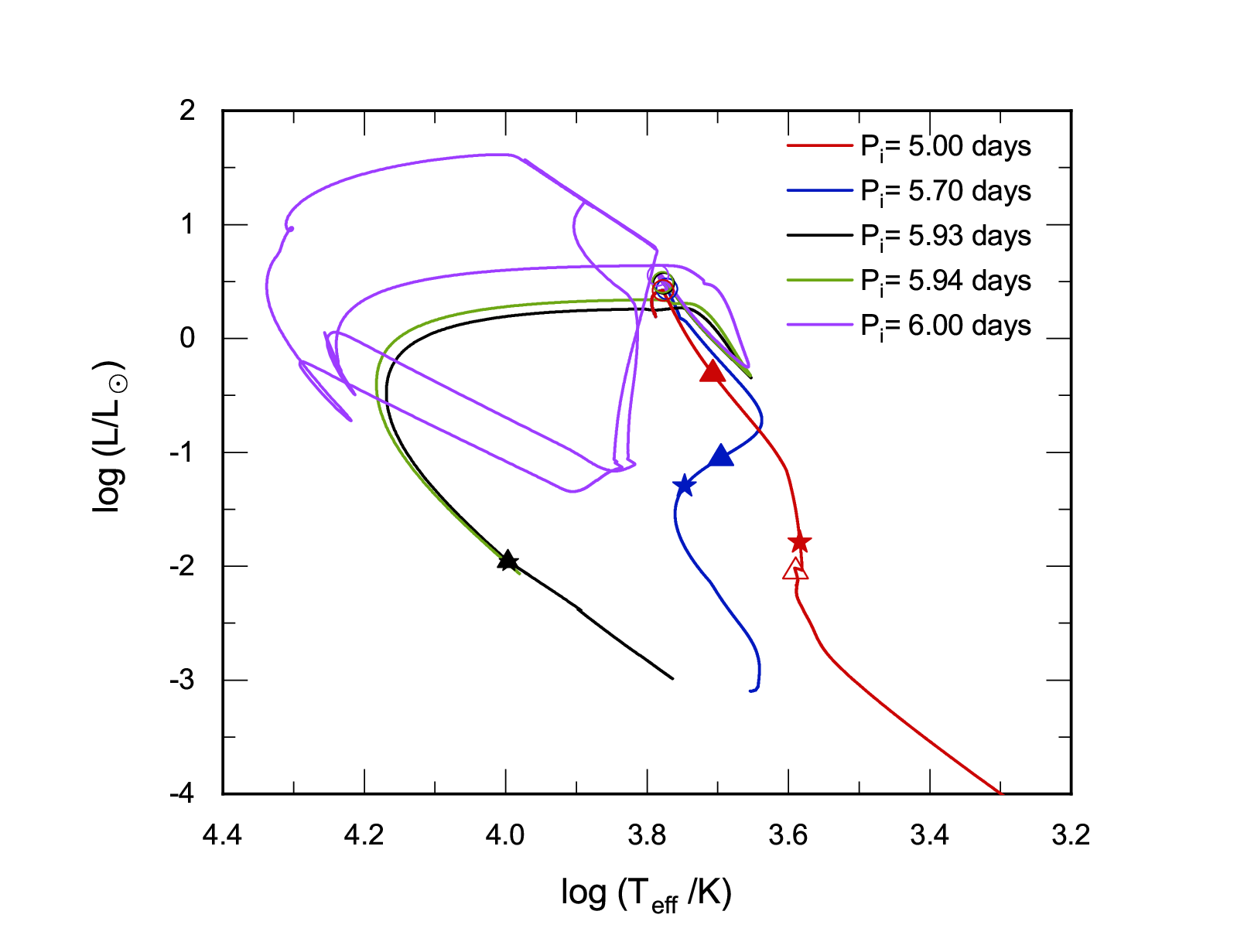}
\caption{Evolution of the companion stars in HR diagram for IMBH-MS binaries in Figures 2 (top panel) and 5 (bottom panel). The open circles, solid triangles, and solid stars indicate the beginning of RLOF, the onset when IMBH binaries are visible by LISA at distances of 1 kpc and 10 kpc, respectively.} \label{fig:HR}
\end{figure}

To better understand the properties and final evolutionary fates of companion stars, Figure \ref{fig:HR} plots the evolution of several IMBH-MS binaries in the HR diagram. In the top panel, companion stars with $M_{\rm d,i}=6.0, 5.7$ and $5.6~M_\odot$ experience four, two, and one hydrogen shell flash, respectively. It is clear that the pre-WD masses are in the range of $0.2-0.32~M_\odot$ (see also Figure \ref{fig:1.80a})  for those three companion stars that experienced hydrogen shell flashes. These are consistent with the pre-WD mass range ($0.212-0.319~M_\odot$) for hydrogen shell flashes in a basic model with $Z=0.02$ \citep[see also table 2 of][]{istr16}. When two systems with $M_{\rm d,i}=6.0, 5.7~M_\odot$ evolve into low-frequency GW sources, the luminosities and effective temperatures are $L\sim 10^{-4}-10^{-3}~L_\odot$ and $T_{\rm eff}\sim5600-7000~\rm K$, respectively, implying that these donor stars have already evolved into WDs. However, the luminosities and effective temperatures of two donor stars with $M_{\rm d,i}=3.0$ and $ 5.5~M_\odot$ are $L\sim 0.1-1~L_\odot$ and $T_{\rm eff}\sim5600~\rm K$ in the low-frequency GW source stage, i.e. they are still in the MS stage experiencing a central hydrogen burning (the abundances of central hydrogen are 0.6 and 0.06 for the donor stars with $M_{\rm d,i}=3.0$ and $ 5.5~M_\odot$, respectively).

In the bottom panel, only one system with $P_{\rm orb,i}=6.0~\rm days$ undergoes two hydrogen shell flashes. Two systems with $P_{\rm orb,i}=5.93$ and $5.94~\rm days$ already evolve into WDs, while their WD masses are slightly less than $0.2~M_\odot$ (see also Figure 5). These two masses are not in the pre-WD mass range ($0.212-0.319~M_\odot$), which can undergo hydrogen shell flashes \citep{istr16}. Hence it is inevitable for a vacancy of hydrogen shell flashes. The luminosities and effective temperatures of two donor stars with $P_{\rm orb,i}=5.0$ and $5.7~\rm days$ are $L\sim (10^{-3}-1)~L_\odot$ and $T_{\rm eff}\sim5000~\rm K$ in the low-frequency GW source stage. Therefore, they are also in the MS stage similar to two systems with $M_{\rm d,i}=3.0$ and $5.5~M_\odot$ in the top panel.

\begin{figure}
\centering
\includegraphics[width=1.15\linewidth,trim={0 0 0 0},clip]{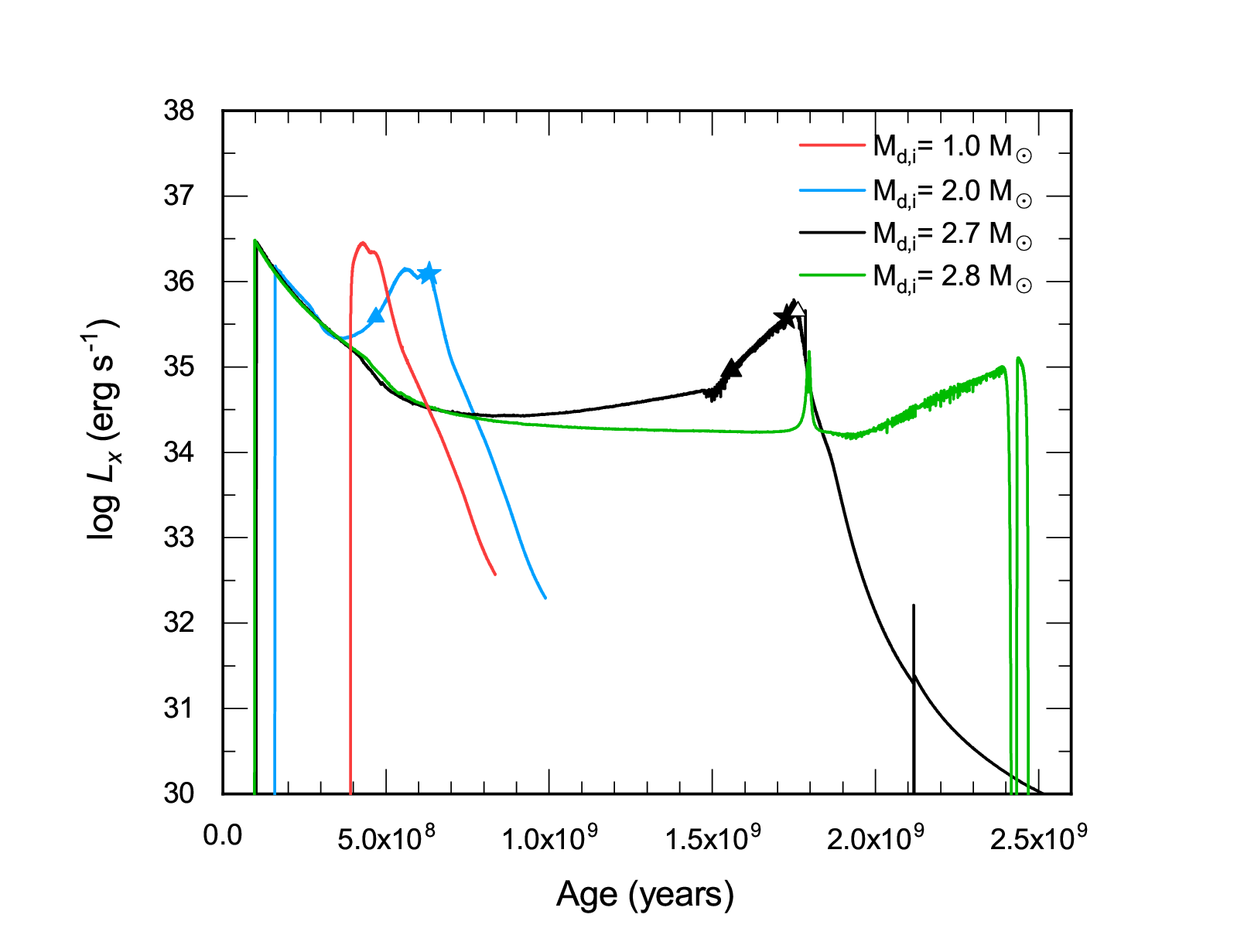}
\includegraphics[width=1.15\linewidth,trim={0 0 0 0},clip]{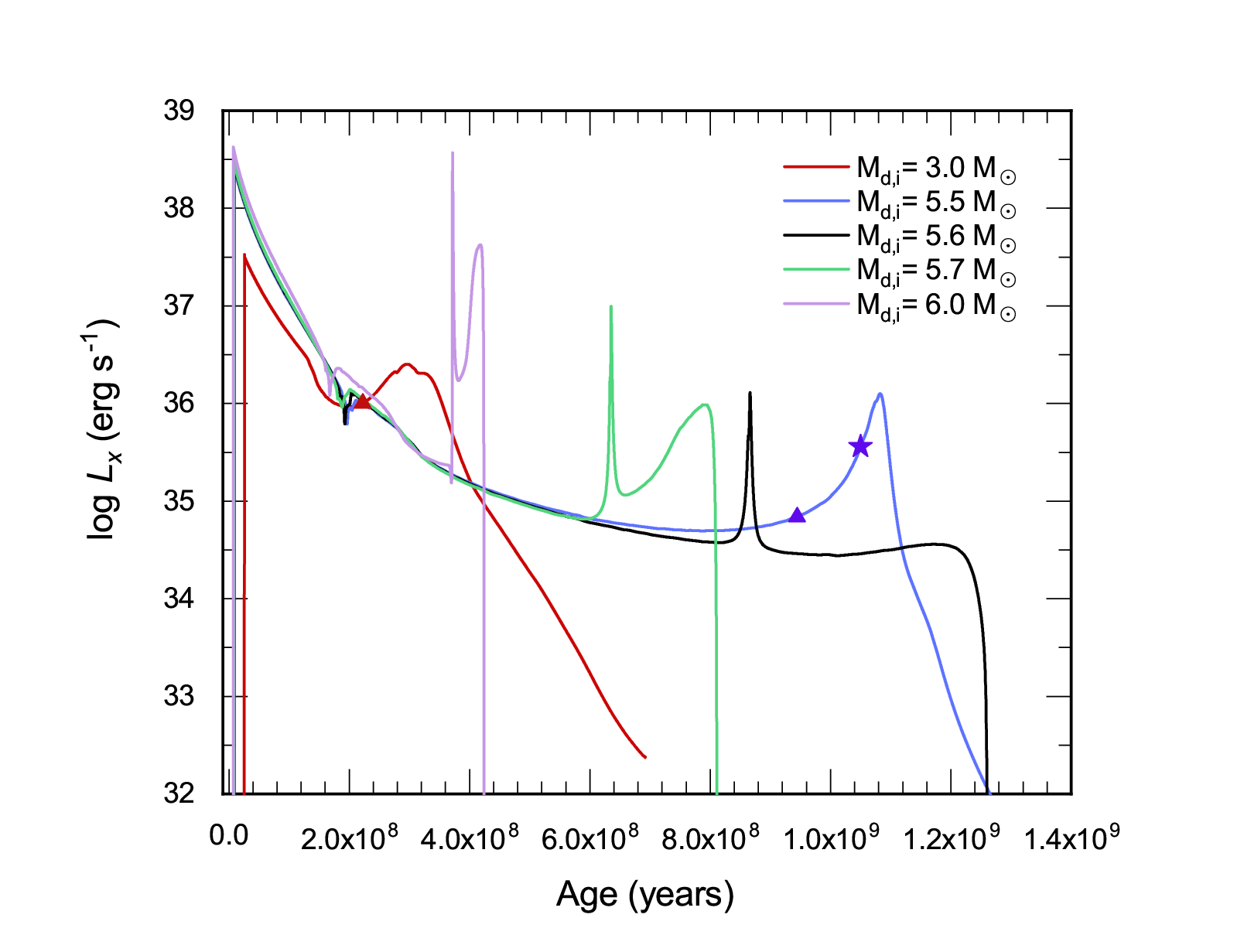}
\caption{Evolution of X-ray luminosity for IMBH X-ray binaries with an initial
orbital period of $P_{\rm orb,i}=1.0~\rm day$ and different initial donor-star masses when $\gamma=1.60$ (top panel, same as Figure \ref{fig:1.60a}) and $\gamma=1.80$ (bottom panel, same as Figure  \ref{fig:1.80a}). The solid triangles and solid stars represent the onsets at which IMBH X-ray binaries are visible by LISA within distances of $d=1$ and 10 kpc, respectively. The first solid triangle in the bottom panel locates the evolutionary curve of $M_{\rm d,i}=3.0~M_{\odot}$.} \label{fig:X_ray}
\end{figure}

\subsection{X-ray luminosity}
When IMBH binaries appear as low-frequency GW sources, they may also be detected as strong X-ray sources. During the accretion of BHs, there exists a high/soft state and a low/hard state, which are dominated by a thermal soft spectrum and a nonthermal hard power law spectrum \citep{nowa95}, respectively. The transition between these two states was believed to take place at a critical accretion rate $M_{\rm crit}$ \citep{Kording2002}. For a high accretion rate $\dot{M}_{\rm acc} >\dot{M}_{\rm crit}$, the X-ray luminosity of the accretion disk is expected to increase with $\dot{M}_{\rm acc}$ just like a standard accretion disk. On the contrary, the X-ray luminosity is proportional to $\dot{M}_{\rm acc}^2$ because of an optically thin advection-dominated accretion flow \citep{nara95}. As a consequence, the X-ray luminosity can be calculated by \citep{Kording2002}
\begin{equation}
    L_{\rm{x}}=
	\begin{cases}
		\epsilon \dot{M}_{\rm acc}c^{2},&\dot{M}_{\rm{crit}}<\dot{M}_{\rm acc} \leq \dot{M}_{\rm{Edd}}\\
		\epsilon \frac{\dot{M}_{\rm acc}^2}{\dot{M}_{\rm{crit}}}c^{2},&\dot{M}_{\rm acc}<\dot{M}_{\rm{crit}},
   \end{cases}
\end{equation}
where $\epsilon=0.1$ is the radiative efficiency of the accretion disk. Similarly to \cite{hopm04}, we take $\dot{M}_{\rm crit}=10^{-7}~M_{\odot}\,\rm yr^{-1}$ for an IMBH with a mass of $1000~M_{\odot}$.

Figure \ref{fig:X_ray} demonstrates the evolution of X-ray luminosities of IMBH X-ray binaries in Figures 1 and 2. Before those IMBH X-ray binaries evolved into low-frequency GW sources, their X-ray luminosities exhibited a relatively low state. Once they appear as low-frequency GW sources, the X-ray luminosities rapidly increase because of a rapid orbital decay. The X-ray luminosities of the IMBH X-ray binaries that are visible for LISA within a distance of 1 and 10 kpc are $\sim 10^{35}-10^{36}~\rm erg\,s^{-1}$. Nonfocusing X-ray telescopes have a detection threshold of $\sim 10^{-11}~\rm  erg\,s^{-1} cm^{-2}$ for X-ray flux \citep{sen21}, corresponding to minimum detectable luminosities of $\sim 10^{33}$ and $\sim 10^{35}~\rm  erg\,s^{-1}$ for distances of 1 and 10 kpc, respectively, which are compatible with our simulated X-ray luminosities. Therefore, compact IMBH X-ray binaries are ideal multi-messenger sources that can be observed in both the low-frequency GW band and electromagnetic wave band. X-ray observations for these low-frequency GW sources would provide a relatively accurate position and distance. However, our simulations indicate that these low-frequency sources cannot appear as ultraluminous X-ray sources with $L_{\rm{x}}>10^{39}~ \rm erg\,s^{-1}$.

\begin{figure}
\centering
\includegraphics[width=1.15\linewidth,trim={0 0 0 0},clip]{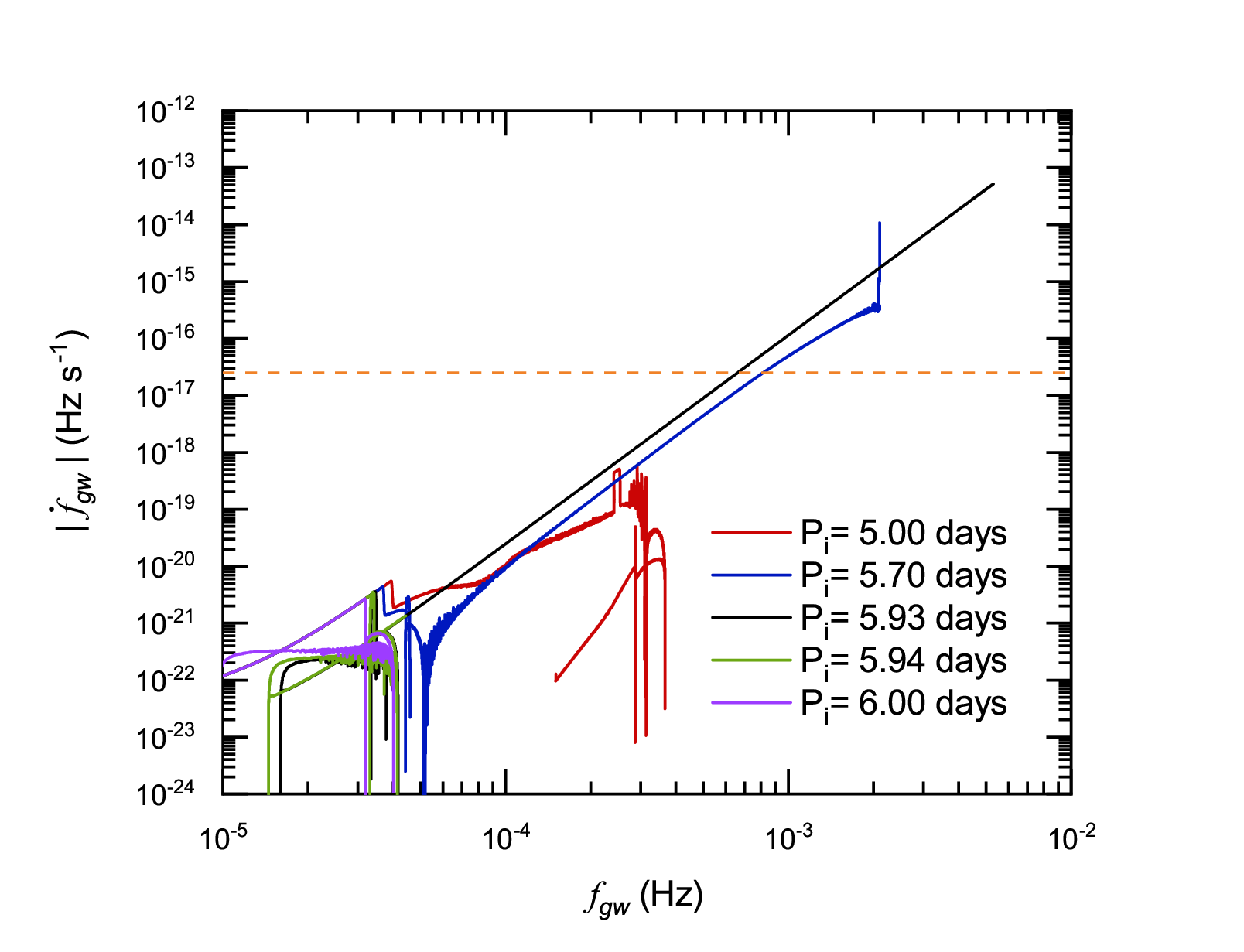}
\caption{Evolution of GW-frequency derivative of IMBH binaries in Figure 6 in the \add{$|\dot{f}_{\rm gw}|$ vs. $f_{\rm gw}$} diagram. The horizontal blue dashed line represents the detection limitation of LISA when $SNR=10$, and $T=4~\rm yr$.} \label{fig:ffdot}
\end{figure}

\subsection{Detectability of chirp signals}
If the GW frequency derivative ($\dot{f}_{\rm gw}$) of a low-frequency GW source can be measured, its chirp mass can be calculated as
\begin{equation}
    \mathcal{M}=\frac{c^{3}}{G}\left(\frac{5}{96\pi^{8/3}}f^{-11/3}_{\rm gw}\dot{f}_{\rm gw}\right)^{3/5}.
\end{equation}
Therefore, it is very important for determining the chirp mass whether $\dot{f}_{\rm gw}$ of a GW source can be accurately measured. The detection limitation of the GW-frequency derivative for LISA is given by \citep{Takahashi2002}
\begin{equation}
    \dot{f}_{\rm gw} \sim 2.5 \times 10^{-17} \left(\frac{10}{SNR}\right)\left(\frac{4~\rm{year}}{T}\right)^2 ~\rm Hz \ s^{-1},
\end{equation}
where $SNR$ is the signal-to-noise ratio of GW signals, $T$ is the total observation time. In principle, LISA can only detect $\dot{f}$ of compact binaries
with a very large SNR, which already evolves to close to the minimum orbital period, and $\dot{f}$ is the largest. Taking a 4-year LISA mission and a GW source with $SNR=10$, the detection limit of LISA is $\dot{f}_{\rm gw,min}\approx2.5\times10^{-17}~\rm Hz\,s^{-1}$.

Figure \ref{fig:ffdot} displays the evolution of the GW frequency derivative of IMBH binaries in Figure 6 in the $\dot{f}_{\rm gw}$ vs. $\dot{f}_{\rm gw}$ diagram. The $\dot{f}$ of two detached systems with $P_{\rm orb,i}=5.93$ and $5.7~\rm days$ can exceed the detection limit of $\dot{f}_{\rm gw,min}\approx2.5\times10^{-17}~\rm Hz\,s^{-1}$ in the frequency range of $0.7-0.9~\rm mHz$. Therefore, it can only accurately measure the GW frequency derivatives and chirp masses for those IMBH binaries whose initial orbital periods are very close to the bifurcation periods. The evolutionary curves of these two systems are titled lines with slope $\Delta \log|\dot{f}_{\rm gw}|/ \Delta \log f_{\rm gw} \approx 4.3$. According to equation (9), pure GW radiation predicts a relation $\dot{f}_{\rm gw}\propto f_{\rm gw}^{11/3}$ for a constant chirp mass in a detached system, resulting in a ratio $\Delta \log|\dot{f}_{\rm gw}|/ \Delta \log f_{\rm gw} = 11/3$, which is less than $4.3$. This implies that the DF of DM also contributes substantially to the orbital decay of these two IMBH binaries. 

\subsection{Birth rate of Galactic IMBH binaries as low-frequency GW sources}
IMBHs could be hosted by globular clusters, young dense clusters, and dwarf galaxies. It is generally thought that, young dense clusters are too young to form compact IMBH X-ray binaries. \cite{holl08} discovered that only 60 globular clusters are able to retain IMBHs with an initial mass of $1000~M_\odot$ in 137 Galactic globular clusters. Assuming the number of IMBHs as $N_{\rm IMBH}=60$, the birth rate of IMBH X-ray binaries as low-frequency GW sources can be expressed as
\begin{equation}
\mathcal{R}=N_{\rm IMBH}\Gamma P t_{\rm GC}/t_{\rm MS},   
\end{equation}
where $\Gamma\sim5\times10^{-8}~\rm yr^{-1}$ is the capture rate that an IMBH captures a companion star \citep{hopm04}, $P$ is the probability that IMBH binaries evolve into LISA sources, $t_{\rm GC}$ and $t_{\rm MS}$ are the lifetimes of globular clusters and MS stars, respectively. In our parameter space, there are about 110 (see also the number of solid stars) IMBH binaries that can evolve toward low-frequency GW sources. In the range of $M_{\rm d,i}=1.0-3.4~ M_{\odot}$ and $P_{\rm orb,i}=0.65-16.82~ \rm days$, there exist 1040 IMBH binaries. If the donor-star masses and orbital periods of those IMBH binaries formed by the tidal capture obey a uniform distribution, the probability is $P\approx0.1$. It seems that this probability is the same as that in \cite{chen20a}. If we adopt a total parameter space as $M_{\rm d,i}=1.0-3.0~ M_{\odot}$ and $P_{\rm orb,i}=0.5-3.5~ \rm days$ as \cite{chen20a}, $P\approx0.2$. Taking $t_{\rm GC}\sim2t_{\rm MS}$, the birth rate of Galactic IMBH binaries as low-frequency GW sources can be estimated to be $\sim1.2\times10^{-6}~\rm yr^{-1}$, which is 2 times higher than ($\sim6\times10^{-7}~\rm yr^{-1}$) in \cite{chen20a}. Therefore, the DF of DM with $\gamma=1.6$ can only enhance the birth rate and widen the parameter space by a factor of $\sim2$.

\subsection{Tidal disruption events}
In a detached IMBH-WD binary, a tidal disruption event will be expected once the WD enters the tidal radius of the IMBH. The radius of a WD is \citep{rapp87,tout97}
\begin{equation}
R_{\rm wd}=0.0115~R_\odot \sqrt{\left(\frac{M_{\rm ch}}{M_{\rm wd}}\right)^{2/3}-\left(\frac{M_{\rm wd}}{M_{\rm ch}}\right)^{2/3}}, 
\end{equation}
where $M_{\rm wd}$ and $M_{\rm ch}=1.44~M_\odot$ are the WD mass and the Chandrasekhar mass limit, respectively. For an IMBH-WD binary with $M_{\rm BH}=1000~M_\odot$ and $M_{\rm wd}=0.3~M_\odot$, the radius of the WD is $R_{\rm wd}=0.018~R_\odot$, and the tidal radius can be calculated as $R_{\rm t}=(M_{\rm BH}/M_{\rm wd})^{1/3}R_{\rm wd}=0.27~R_\odot$. When the tidal radius is equal to the orbital separation, the GW frequency from this system is
\begin{equation}
f_{\rm gw}=\frac{2}{P_{\rm orb}}=\sqrt{\frac{G(M_{\rm BH}+M_{\rm wd})}{\pi^2a^3}}=45~\rm mHz. 
\end{equation}
Therefore, it is possible to simultaneously detect low-frequency GW signals with dozens of mHz and tidal disruption events from an IMBH-WD system.

\subsection{Uncertainties of input parameters in DM spike model}
In the DM spike model, we take a fixed spike radius $r_{\rm sp}=0.2r_{\rm in}$ and a DM background density distribution $\rho_0=\rho_\odot=0.33 \pm 0.03~\rm Gev\,cm^{-3}$. According to Equation (2), larger $r_{\rm sp}$ and higher $\rho_0$ tend to result in higher DM density at the position of the donor star. Higher DM density naturally leads to more efficient loss of angular momentum, making it easier for those IMBH-MS binaries to evolve toward low-frequency GW sources. As a consequence, the initial parameter space that can evolve into low-frequency GW sources is enlarged in the $M_{\rm d,i}-P_{\rm orb,i}$ diagram.

We assume a constant spike index in the detailed stellar evolution model. However, the interaction between the donor star and the DM via gravitational scattering would drive the DM particles into the BH, resulting in a decrease of the spike index by kinetic heating \citep{gned04,merr04}. If gravitational scattering of stars plays an important role, the spike index would evolve to 1.5 due to the kinetic heating process in the heating timescale \citep{merr04,chan2023}
\begin{equation}
\tau_{\rm heat}=3.4\times10^{10}\frac{(M_{\rm BH}/1000~M_\odot)^{1/2}(r_{\rm in}/37.9~\rm pc)^{3/2}}{{\rm ln}\sqrt{M_{\rm BH}/M_{\rm d}}(M_{\rm d}/1~M_\odot)}~\rm yr.  
\end{equation}
When $M_{\rm BH}=1000~M_\odot$, $M_{\rm d,i}=2.7~M_\odot$, $P_{\rm orb,i}=1.0~\rm day$, and $\gamma=1.6$, we have $r_{\rm in}\approx37.9~\rm pc$.
The heating timescales are $15.3$ and $32.8~\rm Gyr$ when the system appears as a low-frequency GW source at a distance of 1 kpc ($M_{\rm d}\sim0.6 ~ M_\odot$) and 10 kpc ($M_{\rm d}\sim0.25 ~ M_\odot$), respectively. These two heating timescales are much longer than the evolutionary timescales ($1.5-1.7~\rm Gyr$, see also Figure 2). Therefore, the change in the spike index can be safely ignored during the evolution of IMBH-MS binaries.

\subsection{A general DF model}
Our work adopts the standard formulation of Chandrasekhar \citep{chan43}, which ignored the contribution of the fast-moving DM particles to the frictional force. According to the standard Chandrasekhar formulation, the DF with any $\gamma>1.5$ tends to circularize the orbits of binary systems \citep{goul03}. Including DF due to the fact that DM particles move faster than an inspiral object, the DF effect would circularize the orbit for any $\gamma>1.8$ \citep{doso24}. Recently, the threshold of the spike index that results in an increase in eccentricity was found to be 2.0 \citep{zhou25}, under which the DF due to fast-moving DM particles will cause the orbit to become more eccentric. However, the increased timescale of eccentricity by the DF is $10^6-10^7~\rm yr$ (see also Figure 2 of \cite{zhou25}), which is much longer than the circularization timescale ($\sim10^4~\rm yr$) due to the tidal interaction between a MS star near RLOF and an IMBH \citep{clar97}. Therefore, the general DF model cannot significantly influence our simulated results.

\section{Conclusions}
In this work, we employ a detailed stellar evolution model to investigate whether the DF of DM can drive IMBH-MS binaries to evolve toward low-frequency GW sources. Our main conclusions are summarized as follows.
\begin{enumerate}
\item The DF of DM provides an efficient mechanism for extracting orbital angular momentum and can drive some IMBH binaries with MS companions to evolve toward low-frequency GW sources. The loss rate of angular momentum is closely related to the donor star mass, the spike index, and the orbital period.
\item Considering the DF of DM, IMBH binaries with low-mass companions tend to evolve into low-frequency GW sources in the X-ray stage. However, some IMBH binaries with high-mass companions may also evolve into low-frequency GW sources because of GR after they become detached systems.
\item A higher spike index produces a higher DM density at the position of the donor star, resulting in a higher loss rate of angular momentum and a shorter evolutionary timescale. When $M_{\rm d,i}=3.0~M_\odot$ and $P_{\rm orb,i}=1.0~\rm days$, there exists a critical spike index $\gamma=1.66$, under which the DF cannot drive IMBH binaries to evolve toward low-frequency GW sources.
\item Those IMBH-MS binaries with initial orbital periods shorter than the bifurcation periods can evolve toward low-frequency GW sources. The low-frequency GW sources whose initial orbital periods are equal to the bifurcation periods are detached systems consisting of an IMBH and a WD, but the donor stars of other GW sources are mass-transferring MS stars.
\item We obtain an initial parameter space of IMBH-MS binaries that can evolve into low-frequency GW sources in the orbital period versus the donor-star mass plane. When $\gamma=1.60$, those IMBH-MS binaries with $M_{\rm d,i}=1.0-3.4~ M_{\odot}$ and $P_{\rm orb,i}=0.65-16.82~ \rm days$ could potentially evolve into visible LISA sources within a distance of $d=10~\rm kpc$. Compared to the standard magnetic braking model, the DF of DM with $\gamma=1.6$ can enhance the birth rate and widen the parameter space by a factor of $\sim2$.

\item In the low-frequency GW source stage, the X-ray luminosities of those IMBH X-ray binaries are $\sim 10^{35}-10^{36}~\rm erg\,s^{-1}$, which are within the detectable range of the X-ray telescopes for a distance of 10 kpc. Only those IMBH X-ray binaries whose initial orbital periods are very close to the bifurcation periods possess a detectable GW frequency derivative and chirp mass.
\end{enumerate}

\begin{acknowledgements}
We thank the referee for a very careful reading and constructive comments that have led to the improvement of the manuscript. We also thank Thomas Tauris, Yi-Ming Hu, Sheng-Hua Yu, and Yan Wang for their helpful discussions. This work was partially supported by the National Natural Science Foundation of China (under grant Nos. 12273014 and 12203051), and the Shandong Province Natural Science Foundation (under grant No. ZR2021MA013).     
\end{acknowledgements}

\bibliographystyle{aa}
\bibliography{ref}

\end{document}